\documentclass[
  aps,
  prfluids,
  reprint,
  onecolumn,
  superscriptaddress,
  amsmath,
  amssymb,
  longbibliography,
  floatfix
]{revtex4-2}

\usepackage[T1]{fontenc}
\usepackage[utf8]{inputenc}
\usepackage{lmodern}
\usepackage{graphicx}
\usepackage{bm}
\usepackage{booktabs}
\usepackage{placeins}
\usepackage{xcolor}
\usepackage[caption=false,labelformat=simple]{subfig}

\usepackage{tikz}
\usetikzlibrary{arrows.meta}
\usepackage[hidelinks]{hyperref}

\definecolor{shockblue}{RGB}{0,142,184}
\definecolor{bodyfill}{RGB}{243,245,246}
\definecolor{bodyline}{RGB}{30,30,30}

\begin{document}

\title{Global modal/non-modal analysis of high-enthalpy Martian entry vehicles}

\author{Adri\'an Ant\'on-\'Alvarez}
\email[Contact author: ]{aantonal@caltech.edu}
\affiliation{Graduate Aerospace Laboratories, California Institute of Technology,
Pasadena, California 91125, USA}

\author{Adri\'an Lozano-Dur\'an}
\affiliation{Graduate Aerospace Laboratories, California Institute of Technology,
Pasadena, California 91125, USA}
\affiliation{Department of Aeronautics and Astronautics,
Massachusetts Institute of Technology, Cambridge, Massachusetts 02139, USA}

\date{\today}

\begin{abstract}
We analyze laminar-to-turbulent transition over a blunt capsule
representative of Martian entry at Mach 26 using global modal and
transient-growth analyses about a thermochemical base flow
characteristic of high-altitude entry conditions.  Across the
Reynolds-number range considered ($Re_\infty = 10^4$--$10^6$), the
axisymmetric global eigenspectrum remains stable, ruling out
temporally unstable axisymmetric global modes as a viable route to
transition. The flow nevertheless supports strong transient growth
concentrated in the shear--entropy layer generated by bow-shock
curvature. Energy-budget analysis reveals a two-stage process:
Reynolds-stress production in the mean shear first amplifies kinetic
disturbances, which then generate entropy fluctuations as they
traverse the strong base-flow entropy gradient. The optimal gain
scales linearly with the Reynolds number based on the capsule nose
radius, whereas boundary-layer-localized disturbances become
competitive at the highest Reynolds numbers examined. These results
identify shear--entropy-layer transient growth as a plausible linear
pathway for seeding transition in high-enthalpy blunt-entry vehicles.
\end{abstract}

\maketitle

\section{Introduction}
\label{sec:introduction}

Reliably predicting the onset of turbulence in hypersonic boundary
layers remains one of the central unresolved problems in high-speed
vehicle design~\cite{schneider2004hypersonic}. Whether the boundary
layer is laminar or turbulent has a direct effect on the dominant
aerodynamic and aerothermal loads, which in turn determine the vehicle
performance envelope and survivability. Once turbulence develops, skin
friction and, most importantly, wall heating increase markedly
relative to laminar conditions~\cite{zoby1981approximate,
  schneider2004hypersonic}. These elevated heat loads place stringent
constraints on the thermal-protection system, with immediate
consequences for structural margins, system mass, and payload
capability. Uncertainty in transition prediction can lead to overly
conservative design margins or underestimated thermal loads.  The Mars
Science Laboratory (MSL) program provides a clear example of how
strongly capsule design margins depend on transition assumptions
through dedicated ground-test campaigns and turbulence-focused
aeroheating analyses~\cite{hollis2005transition, hollis2007turbulent,
  hollis2010blunt, wright2006modeling}. Post-flight reconstructions
for MSL and Mars 2020/Perseverance reinforced the same concern by
documenting the realized aerothermal environment, including evidence
of boundary-layer transition and turbulent conditions in
flight~\cite{bose2014reconstruction, edquist2022mars,
  alpert2022inverse}.

In this work, we investigate the linear amplification mechanisms that
may drive laminar-to-turbulent transition over hypersonic blunt bodies
under high-enthalpy Mars-entry conditions. To this end, we perform
global modal and non-modal stability analyses of a representative
capsule geometry at high-altitude Martian entry conditions. Below, we
review the relevant background and identify the specific knowledge
gaps addressed by the present study.

\subsection{Modal stability}
\label{sec:modal-stability}

The classical framework for understanding transition in compressible
boundary layers is rooted in modal stability theory. Early analyses
considered infinitesimal perturbations to wall-parallel base flows
governed by the linearized compressible Navier--Stokes
equations~\cite{lees1946investigation}, establishing the compressible
analogue of Rayleigh's inflection-point criterion for inviscid
instability. When viscosity is retained, compressible boundary layers
support first-mode disturbances that are analogous to the
Tollmien--Schlichting waves of incompressible boundary
layers~\cite{lin1945stability, mack1984boundary,
  fedorov2011transition}. These first-mode disturbances extract energy
from the mean shear through Reynolds-stress work and dominate
transition for moderate freestream Mach numbers. A major departure
from incompressible behavior was identified by
\citet{mack1984boundary}, who showed that compressible boundary layers
admit additional discrete unstable modes with no incompressible
analogue: the Mack, or second, modes. These modes arise from acoustic
trapping and resonance within the boundary layer and become
increasingly important as the freestream Mach number increases.

For hypersonic vehicles, the relevant modal mechanism depends strongly
on the post-shock state. In blunt-body configurations such as Entry,
Descent, and Landing (EDL) vehicles, strong detached bow shocks
decelerate the outer flow to subsonic or weakly supersonic conditions
($M \lesssim 2$), favoring first-mode dynamics. By contrast,
sharp-nosed configurations preserve larger post-shock Mach numbers and
are therefore more susceptible to second-mode
instabilities~\cite{fedorov2011transition}.

Flight data from the HIFiRE program provide particularly valuable
evidence of transition on slender high-speed configurations. HIFiRE-1
documented boundary-layer transition on a slender cone, whereas
HIFiRE-5b documented transition over a three-dimensional elliptic cone
for which multiple mechanisms, including crossflow-related effects,
were implicated~\cite{kimmel2015hifire,juliano2018hifire}.  These
observations are consistent with the broader literature linking
cold-wall slender configurations to second-mode
dynamics~\cite{fedorov2011transition}. In particular, the
measurements reported by \citet{juliano2018hifire} revealed a
multilobed transition front, suggesting that different instability
mechanisms were active in different regions of the elliptic cone.

Later developments relaxed the parallel-flow approximation in order to
account for the slow streamwise evolution of realistic boundary
layers. The parabolized stability equations
(PSE)~\cite{bertolotti1991analysis,herbert1997parabolized} provide
one such framework by marching disturbances downstream while filtering
upstream-propagating components. Nonlinear extensions, commonly
referred to as NPSE, capture modal interactions and permit the
analysis of secondary-instability mechanisms and resonance
processes~\cite{bertolotti1992linear,chang1993linear,chang1994oblique}. More
recently, one-way spatial integration of the Navier--Stokes equations
(OWNS) has extended the class of weakly non-parallel flows that can be
analyzed within a marching framework~\cite{TowneColonius2015JCP,
  TowneRigasPickeringColonius2022JFM,
  SleemanLakebrinkColonius2025AIAAJ}. Together, these approaches have
shown that non-parallel effects can substantially modify disturbance
growth and improve transition predictions in spatially developing
high-speed boundary layers (see, e.g.,
Refs.~\cite{herbert1997parabolized, Paredes2019, Paredes2022}).

Even so, streamwise-marching approaches inherently rely on a slowly
varying streamwise evolution and are therefore less suitable for flows
with rapid spatial variations or strong global coupling, as commonly
encountered in hypersonic shock layers over blunt vehicles. Such flows
motivate the use of global stability
analysis~\cite{theofilis2003advances,theofilis2011global}, which
formulates the disturbance dynamics as a space-filling eigenvalue
problem without imposing a downstream-marching assumption. This
framework has revealed instability mechanisms in complex high-speed
configurations that are not captured by local or weakly non-parallel
approaches. For example, \citet{paredes2016linear} identified multiple
unstable global modes over a hypersonic elliptic cone. Similarly,
\citet{hildebrand2018simulation} used global stability analysis to
identify a three-dimensional stationary instability in a Mach~5.92
oblique shock-wave/boundary-layer interaction, associated with the
strongly coupled dynamics of the shock-induced separation region.

\subsection{Non-modal analysis}
\label{sec:nonmodal-analysis}

Another important route to transition is non-modal disturbance
growth~\cite{reshotko2001transient}. In hypersonic flows, this
perspective has proved especially valuable in explaining discrepancies
between experimental observations and predictions based solely on
exponentially growing eigenmodes. Transition often occurs earlier than
predicted by modal theory~\cite{schneider2004hypersonic}, and
significant amplification can be measured in regions where the
eigenspectrum is stable or only weakly unstable. Non-normal linear
operators resolve this apparent contradiction: over finite times, they
can produce transient energy amplification far larger than that
associated with any single eigenmode~\cite{schmid2007nonmodal}.

Two classical mechanisms dominate the incompressible transient-growth
literature. The lift-up effect converts streamwise vortices into
elongated high- and low-speed
streaks~\cite{landahl1980note,butler1992three}, while the Orr
mechanism amplifies tilted disturbances as mean shear reorients them
before eventual decay~\cite{orr1907stability}.  Both mechanisms also
appear in compressible and hypersonic boundary
layers~\cite{hanifi1996transient,Paredes2016}. The effect of Mach
number on transient growth depends on the wall thermal condition: in
cooled flat-plate boundary layers, increasing the Mach number can
reduce the optimal gain~\cite{Bitter2014}.  In high-speed
configurations, the Orr mechanism is particularly relevant because it
helps explain the amplification of oblique structures during the
premodal stage of transition (see, e.g., Refs.~\cite{Bitter2014,
  Paredes2016}). Related non-modal mechanisms have also been identified
in hypersonic shock-wave/boundary-layer
interactions~\cite{Dwivedi2020}.

The growth of computational capabilities has also made fully global
non-modal analyses of realistic hypersonic configurations
feasible. These studies reveal transient-growth mechanisms that are
not captured by local parallel or weakly non-parallel models (see,
e.g., Ref.~\cite{quintanilha2022transient}). A representative
example is the elliptic-cone study of
\citet{quintanilha2022transient}, who reported optimal gains as large
as $O(10^4)$ and found strong sensitivity of the gain scaling to
flight condition. Their results reinforced the idea that substantial
linear amplification may occur even when modal instabilities are
absent or weak.

Particularly relevant to the present work is the shear--entropy-layer
mechanism that arises in hypersonic blunt-body
flows. \citet{hornung2001shock} first identified this behavior
experimentally in spherical geometries and linked it to the entropy
layer generated by curved strong shocks. The same mechanism is now
suspected to be a central ingredient in the transition-reversal
paradox on blunt forebodies: increasing nose bluntness stabilizes
conventional boundary-layer modes, yet experiments show that
sufficiently blunt noses can nevertheless promote earlier
transition~\cite{marineau2014mach,jewell2017boundary,paredes2019nose}. Non-modal
analyses explain this behavior by showing that the detached entropy
layer forms a shear region capable of large transient growth even when
the wall boundary layer remains modally
stable~\cite{paredes2017blunt,paredes2019nose,paredes2020mechanism}. Subsequent
nonlinear simulations further demonstrated that disturbances amplified
in the entropy layer can couple with the wall boundary layer and
accelerate breakdown~\cite{scholten2024nonlinear}.

\subsection{Current challenges and limitations}
\label{sec:challenges}

The literature reviewed above exposes several gaps that are
especially relevant to blunt entry capsules operating under
flight-representative conditions. First, much of the modern
hypersonic-transition literature has focused on cones, elliptic
cones, and related slender forebodies, for which the dominant linear
physics is usually sought in Mack's second mode or in
crossflow-related mechanisms. These mechanisms rely on a
boundary-layer-edge Mach number that remains large behind the
attached or weakly detached shock of a slender body. A capsule
produces a fundamentally different flow topology. The strong detached
bow shock decelerates the flow to subsonic or low-supersonic
conditions over most of the forebody, so the edge Mach number never
reaches the range in which the second mode operates and the
instability that dominates the slender-body literature is simply
unavailable. What remains is strong global non-parallelism, rapid
thermodynamic variation through the shock layer, and an extended
forebody entropy--shear layer generated by shock curvature. The
relevant amplification mechanisms therefore need not resemble those
of canonical slender configurations, which motivates a fully global
analysis tailored to EDL-vehicle flows.

Second, a substantial fraction of the available computational and
experimental literature has been developed for low-enthalpy flows
modeled as calorically perfect gases. Actual Mars-entry conditions,
however, involve high enthalpy, active thermochemistry, and freestream
Mach numbers as large as $M_\infty \approx 30$. These effects modify
both the base flow and the disturbance dynamics, altering shock-layer
structure, thermodynamic gradients, acoustic propagation, and modal
coupling. Assessing their influence requires a formulation that is
thermochemically consistent with the entry regime of interest.

The present work leverages global modal and optimal transient-growth
analyses of a high-enthalpy axisymmetric base flow representative of
Mars-entry conditions. Our objective is to identify the dominant
linear amplification pathway and determine whether it arises from
temporally unstable global modes, disturbances localized within the
wall boundary layer, or transient growth associated with the
entropy--shear layer. 
The companion study by \citet{AntonAlvarez2026bow} examines the
receptivity of high-enthalpy blunt-entry flows to sustained freestream
disturbances, whereas the present global modal and transient-growth
analyses isolate the underlying linear amplification mechanisms in
the absence of such forcing.
The remainder of the paper is organized as
follows. Section~\ref{sec:methods} describes the physical model,
numerical framework, and linear-stability
methodology. Section~\ref{sec:modal-results} presents the global
modal results. Section~\ref{sec:transient-results} analyzes the
optimal transient growth and its energy
budgets. Section~\ref{sec:reynolds-dependence} then characterizes how
the dominant amplification mechanisms vary with Reynolds number.

\section{Methods}
\label{sec:methods}

\subsection{Problem setup}
\label{sec:problem-setup}

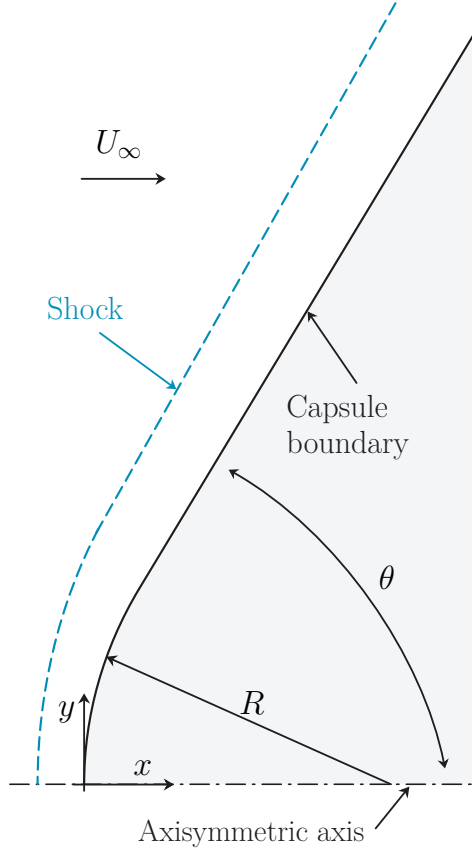
\begin{figure}[tbp]
\centering
\resizebox{0.34\linewidth}{!}{%
\begin{tikzpicture}[
  x=1pt,y=-1pt,
  line cap=round,
  line join=round,
  >=Stealth,
  body/.style={draw=bodyline,line width=1.35pt},
  shock/.style={draw=shockblue,line width=1.35pt,dash pattern=on 8pt off 4.4pt},
  axisline/.style={draw=bodyline,line width=1.0pt,dash pattern=on 12pt off 4.5pt on 1.5pt off 4.5pt},
  guide/.style={draw=bodyline,line width=1.15pt},
  guideblue/.style={draw=shockblue,line width=1.15pt},
  ann/.style={font=\fontsize{20}{22}\selectfont},
  annsmall/.style={font=\fontsize{19}{21}\selectfont},
  mathann/.style={font=\fontsize{21}{21}\selectfont},
  every node/.style={inner sep=0pt}
]

\path[use as bounding box] (0,0) rectangle (288,540);
\clip (0,0) rectangle (288,540);

\fill[bodyfill]
  (80.722656,367.570312)
  -- (291.089844,18.660156)
  -- (291.089844,489.773438)
  -- (46.375000,489.773438)
  .. controls (46.375000,447.121094) and (58.175781,405.128906) ..
  (80.722656,367.570312)
  -- cycle;

\draw[body]
  (80.722656,367.570312)
  .. controls (58.175781,405.128906) and (46.375000,447.121094) ..
  (46.375000,489.773438);

\draw[body]
  (80.722656,367.570312) -- (291.089844,18.660156);

\draw[axisline]
  (0,489.773438) -- (295,489.773438);

\draw[shock]
  (55.406250,329.949219)
  .. controls (30.609375,376.839844) and (17.316406,432.621094) ..
  (17.316406,489.773438);

\draw[shock]
  (54.035156,332.441406) -- (241.621094,0);

\draw[guide,-{Stealth[length=6.5pt,width=7.5pt]}]
  (238.0,489.8) -- (60.144531,410.050781);

\node[mathann] at (151.0,438.5) {$R$};

\draw[guide,{Stealth[length=5.8pt,width=7pt]}-{Stealth[length=5.8pt,width=7pt]}]
  (136.6,293.1)
  .. controls (205,332) and (258,410) ..
  (271.7,479.9);

\node[mathann] at (235.5,360.5) {$\theta$};

\draw[guide,-{Stealth[length=6.5pt,width=7.5pt]}]
  (39.589844,489.941406) -- (102.558594,489.941406);

\draw[guide,-{Stealth[length=6.5pt,width=7.5pt]}]
  (46.375000,493.910156) -- (46.375000,432.898438);

\node[mathann] at (83.0,479.5) {$x$};
\node[mathann] at (35.5,444.4) {$y$};

\node[mathann] at (68.0,89.3) {$U_\infty$};

\draw[guide,-{Stealth[length=6.5pt,width=7.5pt]}]
  (44.769531,111.269531) -- (97.382812,111.269531);

\node[shockblue,ann,anchor=west] at (23.2,192.0) {Shock};

\draw[guideblue,-{Stealth[length=6pt,width=7pt]}]
  (56.2,207.6) -- (104.0,242.85);

\node[bodyline,ann,anchor=west,align=left] at (172.6,266.5) {Capsule\\boundary};

\draw[guide,-{Stealth[length=6pt,width=7pt]}]
  (216.5,238.9) -- (185.9,194.15);

\node[bodyline,annsmall,anchor=west] at (80,520.3) {Axisymmetric axis};

\draw[guide,-{Stealth[length=6.5pt,width=7.5pt]}]
  (228.4,517.4) -- (249.5,491.2);

\end{tikzpicture}%
}
\caption{Reference coordinate system and geometric notation for the simplified capsule.}
\label{fig:setup}
\end{figure}

We consider the flow over an axisymmetric simplified capsule
representative of EDL vehicles. The geometry consists of a spherical
nose cap of radius $R$ blending smoothly into a frustum of half-angle
$\theta = 60^{\circ}$, as shown in Fig.~\ref{fig:setup}.  The
junction between the spherical and conical portions is tangent.
A Cartesian reference frame is adopted with the $x$ axis aligned with
the freestream direction and with the axis of symmetry of the
body. The origin is located at the stagnation point, which coincides
with the apex of the spherical nose. The base flow is assumed
axisymmetric.

The freestream is characterized by density $\rho_\infty$, temperature
$T_\infty$, and velocity magnitude $U_\infty$. The conditions used in
the present study are listed in Table~\ref{tab:freestream}. They were
selected to represent a point along the MSL and Perseverance
trajectories for which reconstructed aerothermal data indicate the
onset of enhanced heating associated with
transition~\cite{bose2014reconstruction, edquist2022mars,
  alpert2022inverse}. Under these conditions, a detached bow shock
forms upstream of the capsule nose. The resulting post-shock state
determines the flow conditions that govern the stability
behavior investigated here.
\begin{table}[t]
\caption{Freestream conditions used to compute the base flow. The mole
  fractions $X_i$ denote the composition of the Martian atmosphere.}
\label{tab:freestream}
\centering
\small
\setlength{\tabcolsep}{4pt}
\begin{ruledtabular}
\begin{tabular}{@{}cccp{0.50\textwidth}c@{}}
$U_\infty$ (m/s) &
$\rho_\infty$ (kg/m$^3$) &
$T_\infty$ (K) &
Freestream composition &
$M_\infty$ \\
\colrule
5260 &
0.000822 &
177 &
$X_{\text{CO}_2}:0.9556$, $X_{\text{N}_2}:0.0270$,
$X_{\text{Ar}}:0.0160$, $X_{\text{O}_2}:0.0014$ &
26.1
\end{tabular}
\end{ruledtabular}
\end{table}

\subsection{Modeling equations}
\label{sec:modeling-equations}

\subsubsection{Compressible Navier--Stokes equations}
\label{sec:ns}

The flow is modeled as a continuum, viscous, heat-conducting,
single-temperature equilibrium gas mixture. Unless otherwise stated,
length, velocity, density, pressure, and time are non-dimensionalized by
$R$, $U_\infty$, $\rho_\infty$, $\rho_\infty U_\infty^2$ and
$R/U_\infty$, respectively. The specific internal energy is
non-dimensionalized by $U_\infty^2$, the temperature by
$U_\infty^2/c_{v\infty}$, and the entropy by $c_{v\infty}$. With these
conventions, the governing equations are
\begin{subequations}
\label{eq:ns}
\begin{align}
\frac{\partial \rho}{\partial t}
+\frac{\partial(\rho u_i)}{\partial x_i}
&=0,
\label{eq:ns-a}\\[3pt]
\frac{\partial(\rho u_i)}{\partial t}
+\frac{\partial(\rho u_i u_j)}{\partial x_j}
&=
-\frac{\partial\bigl[(\gamma^* - 1)\rho e\bigr]}{\partial x_i}
+\frac{\partial\tau_{ij}}{\partial x_j},
\label{eq:ns-b}\\[3pt]
\frac{\partial \left(\frac{1}{2}\rho u_i u_i + \rho e\right)}{\partial t}
+ \frac{\partial \left[\left(\frac{1}{2}\rho u_i u_i + \gamma^*\rho e\right)u_j\right]}{\partial x_j}
&= \frac{\partial (\tau_{ij}u_i)}{\partial x_j} - \frac{\partial q_i}{\partial x_i},
\label{eq:ns-c}\\[3pt]
\tau_{ij}
&=
\frac{\mu^*}{Re_\infty}
\left(
\frac{\partial u_j}{\partial x_i}
+\frac{\partial u_i}{\partial x_j}
-\frac{2}{3}
\frac{\partial u_k}{\partial x_k}\delta_{ij}
\right),
\label{eq:ns-d}\\[3pt]
q_i
&=
-\frac{\gamma_\infty k^*}{Re_\infty Pr_\infty}
\frac{\partial}{\partial x_i}
\left(\frac{e}{c_v^*}\right).
\label{eq:ns-e}
\end{align}
\end{subequations}
Here $u_i$ are the velocity components, $\rho$ is the density,
$e$ is the specific internal energy, $p$ is the pressure, $T$ is
the temperature, $\tau_{ij}$ is the viscous stress and $q_i$ is
the heat flux. The effective thermodynamic quantities are
$\gamma^* = 1 + p/(\rho e)$ and
$c_v^* = e/T$, while the non-dimensional transport
coefficients are $\mu^*=\mu/\mu_\infty$ and
$k^*=k/k_\infty$. The corresponding non-dimensional groups are
\begin{equation}
Re_\infty
=
\frac{\rho_\infty U_\infty R}{\mu_\infty},
\qquad
\gamma_\infty
=
\frac{c_{p\infty}}{c_{v\infty}},
\qquad
Pr_\infty
=
\frac{\mu_\infty c_{p\infty}}{k_\infty},
\label{eq:nondim_groups}
\end{equation}
where $c_{p\infty}$ is the freestream specific heat at constant
pressure. The thermochemical models for $\gamma^*$ and $c_v^*$ along
with the transport models for $k^*$ and $\mu^*$ are discussed in the
next section. The governing equations are closed with the
ideal-mixture equation of state
\[
p=\rho R_g T,
\]
where the local mixture gas constant $R_g$ is non-dimensionalized by
$c_{v\infty}$. 

The use of the continuum equations is justified by the small
body-scale Knudsen number:
\[
Kn_\infty
\simeq
\sqrt{\frac{\pi\gamma_\infty}{2}}\,
\frac{M_\infty}{Re_\infty}.
\]
At the freestream state in Table~\ref{tab:freestream}, this estimate
gives $Kn_\infty\simeq 4\times10^{-4}$ for
$Re_\infty=10^5$. Additionally, the local Knudsen number is even
smaller in the compressed post-shock region. Therefore, the flow is
well within the continuum regime at the body scale relevant to the
present analyses.

\subsubsection{Thermochemical and transport models}
\label{sec:thermochem}

We assume a chemically and thermally equilibrated ideal-gas mixture
for both the base-flow calculations and the modal and transient-growth
analyses.  Under this assumption, separate species-transport and
vibrational-energy equations are not solved. Instead, the local
composition is obtained algebraically from the equilibrium
relations. The constitutive models for the freestream molar
composition $\boldsymbol{X}_\infty$ 
\[
\gamma^*=\gamma^*(\rho,e;\boldsymbol{X}_\infty), \
c_v^*=c_v^*(\rho,e;\boldsymbol{X}_\infty), \
\mu^*=\mu^*(\rho,e;\boldsymbol{X}_\infty), \
k^*=k^*(\rho,e;\boldsymbol{X}_\infty)
\]
are described in Ref.~\cite{AntonAlvarez2026HYMOR}. The equilibrium
thermochemical model also provides $R_g$, the local equilibrium
composition, and the local speed of sound.

For reference, we define the
chemical and vibrational Damk\"{o}hler numbers using the body-scale
post-shock flow time $ \tau_{\mathrm{flow}} = R/U_2^{\mathrm{RH}}$,
where $U_2^{\mathrm{RH}}$ is the post-shock normal velocity from the
corresponding normal-shock Rankine--Hugoniot estimate. Thus
\begin{equation}
Da_{\mathrm{chem}}
=
\frac{\tau_{\mathrm{flow}}}{\tau_{\mathrm{chem}}},
\qquad
Da_{\mathrm{vib}}
=
\frac{\tau_{\mathrm{flow}}}{\tau_{\mathrm{vib}}}.
\label{eq:damkohler_methods}
\end{equation}
For the baseline Mars-entry state at $Re_\infty = 100\,000$,
Appendix~\ref{sec:equilibrium-validity} gives
$Da_{\mathrm{chem}}=O(10^2)$--$O(10^3)$ and
$Da_{\mathrm{vib}}=O(10^4)$--$O(10^5)$ in the critical
velocity--altitude region. The relaxation times are therefore short
relative to the residence time of the gas in the forebody shock layer,
supporting the use of an equilibrium model for the stability
calculations reported here. The equilibrium calculations and
chemical-relaxation-time estimates are obtained with
Cantera~\cite{cantera}, while the vibrational-relaxation-time
estimates use the Millikan--White
correlation~\cite{millikan1963systematics}, as described in
Appendix~\ref{sec:equilibrium-validity}.

\subsubsection{Boundary conditions}
\label{sec:boundary-conditions}

Boundary conditions are imposed at the bow shock, the wall, the
symmetry axis, and the downstream outflow plane:
\begin{itemize}
\item \emph{Shock compatibility.}  The computational domain includes
  only the post-shock region.  At the fitted bow shock, the downstream
  state satisfies the Rankine–Hugoniot jump relations with the uniform
  freestream, closed by the equilibrium-mixture thermodynamic
  model. In the linearized calculations, the jump relations and shock
  kinematic condition are linearized consistently about the
  fitted-shock base state.
\item \emph{Wall.} No-slip ($\bm{u}=\bm{0}$) and adiabatic-wall
  ($q_w=0$) conditions are imposed at the capsule surface. Although
  flight vehicles are often more accurately represented by cold-wall
  conditions, cases with cooled isothermal walls were also examined
  and found not to affect the shear--entropy-layer mechanisms
  identified here, which are localized far from the wall.
\item \emph{Axis of symmetry.} Standard axisymmetric boundary
  conditions are enforced along the geometric symmetry axis.
\item \emph{Outflow.} Non-reflecting characteristic conditions are
  applied at the downstream boundary following
  \citet{poinsot1992boundary}.
\end{itemize}

\subsection{Numerical solver}
\label{sec:discretization}

The results reported here are obtained with HYMOR, an open-source code
for global stability analysis in high-enthalpy hypersonic flows.
The implementation details, verification cases, and usage examples
are described in Ref.~\cite{AntonAlvarez2026HYMOR}. Here, we only summarize
the ingredients needed to interpret the present results.

HYMOR solves the axisymmetric compressible Navier--Stokes equations
with a second-order finite-volume discretization on a curvilinear
body-fitted grid. Face fluxes are obtained by linear interpolation to
the cell faces and midpoint quadrature over each face. Time
advancement is performed with an explicit fourth-order Runge--Kutta
scheme.

A shock-fitting strategy is used so that the bow shock remains a sharp
discontinuity in both the base-flow and linearized calculations. This
avoids shock-capturing artifacts such as the carbuncle
phenomenon~\cite{pandolfi2001numerical} and, more importantly for the
present work, preserves the correct linear interaction between
infinitesimal disturbances and the shock. In the stability analysis,
the shock compatibility conditions are linearized consistently with
the fitted-shock formulation, so shock motion is included as part of
the post-shock disturbance dynamics.  The fitted shock is represented
by a cubic-spline curve that evolves independently of the flow
grid. As such, the shock location is not constrained to grid nodes,
which provides sub-grid accuracy, avoids carbuncle-type artifacts, and
preserves the nominal spatial order of accuracy.

The calculations were carried out on three mesh refinement levels,
denoted coarse, base, and fine, whose resolution was scaled with the
Reynolds number.  For the representative case $Re_\infty = 100\,000$,
the coarse, base, and fine grids contain $800 \times 120$,
$1200 \times 200$, and $1800 \times 300$ cells, respectively.  A
three-level refinement study showed that the least-stable eigenvalue
and the peak transient energy gain vary by less than $2\%$ between the
base and fine grids; the base grid is therefore adopted for all
results reported hereafter.

\subsection{Linear stability analysis}
\label{sec:linear-stability}

We investigate linear amplification using global modal analysis and
non-modal transient-growth analysis.  The conservative flow-state
vector $q$ comprises the density $\rho$, the momentum densities $\rho
u_i$, and the total energy density $\rho e+\tfrac{1}{2}\rho u_i
u_i$. Continuous quantities are denoted by non-bold symbols, whereas
their semi-discrete counterparts are denoted in bold. In particular,
$\bm{q}$ collects the conservative variables over the post-shock
finite-volume cells. The total state is decomposed as $q = q_0 + q'$,
where the subscript $0$ denotes the steady base flow and the prime
denotes an infinitesimal disturbance. In the present work,
disturbances are restricted to the axisymmetric class. The conclusions
below therefore concern axisymmetric global modes and axisymmetric
optimal disturbances only.

Because the bow shock is fitted, the discrete state contains both the
post-shock flow field and the displacement of the fitted shock. We
therefore linearize about the extended shock-fitted base state and
write the extended disturbance as
\begin{equation}
\bm{q}'_e
=
\begin{bmatrix}
\bm{q}'\\
\bm{\eta}'
\end{bmatrix},
\label{eq:augmented_state_stability}
\end{equation}
where $\bm{q}'$ contains the cell-centered conservative perturbation
variables in the post-shock region and $\bm{\eta}'$ contains the
perturbations of the shock-spline degrees of freedom, i.e., the normal
displacement of the fitted shock about its steady position. The
shock-displacement coordinate $\bm{\eta}'$ enters the dynamics through
the Rankine--Hugoniot compatibility and shock kinematic conditions,
so that disturbances of the post-shock flow and motion of the fitted
shock evolve as a single coupled system.

Linearization about the steady extended base state $\bm{q}_{e,0}$,
obtained from a direct numerical simulation, yields the semi-discrete
system
\begin{equation}
\frac{\partial \bm{q}'_e}{\partial t} = \bm{L}(\bm{q}_{e,0})\,\bm{q}'_e,
\qquad
\bm{L}(\bm{q}_{e,0})
\equiv
\left.
\frac{\delta \bm{N}}{\delta \bm{q}_e}
\right|_{\bm{q}_{e,0}},
\label{eq:linearized-system}
\end{equation}
where $\bm{N}$ is the nonlinear shock-fitted residual, including the
interior finite-volume equations, the shock kinematic condition and
the Rankine--Hugoniot compatibility conditions, and $\bm{L}$ is the
corresponding Jacobian evaluated at $\bm{q}_{e,0}$ and assembled by
finite differences. The matrix $\bm{L}(\bm{q}_{e,0})$ is written
simply as $\bm{L}$ hereafter. Additional implementation details are
given by \citet{AntonAlvarez2026HYMOR}.

For the modal analysis, we solve the eigenvalue problem
$\lambda\bm{q}'_e = \bm{L}\bm{q}'_e$. Only post-shock perturbations
are considered. Because the uniform freestream is both modally stable
and supersonic, it is dynamically decoupled from the downstream
eigenspectrum; any exponentially growing mode must therefore reside
within the post-shock region.  The eigenpairs are computed with a
GPU-based iterative algorithm that combines spectral mapping with
Arnoldi iteration.

For the transient-growth analysis, the freestream is again assumed
unperturbed and only post-shock disturbances are retained.
Disturbance energy is measured with Chu's norm~\cite{chu1965energy},
\begin{equation}
E = \int_{V_D}\left[
\underbrace{\frac{\rho_0 a_0^2}{2(\gamma_0^*p_0)^2}p'^2}_{\text{pressure}}
+ \underbrace{\frac{\rho_0}{2}u_i'u_i'}_{\text{kinetic}}
+ \underbrace{\frac{(\gamma_0^*-1)p_0}{2\gamma_0^*}\left(\frac{s'}{R_{g,0}}\right)^2}_{\text{entropic}}
\right]dV,
\label{eq:chu}
\end{equation}
where $V_D$ is the flow volume downstream of the bow shock, $a_0$ is
the base-flow speed of sound, $R_{g,0}$ is the local base-flow mixture gas
constant, and $\gamma_0^*$ is the effective ratio of specific heats of
the base flow. The three contributions in Eq.~\eqref{eq:chu} correspond to
pressure, kinetic, and entropic disturbance energy.

The discrete gain is defined as $G(t)=E(t)/E(0)$ over the post-shock
finite-volume cells. The initial disturbance is restricted to the
post-shock flow field, with zero initial displacement of the fitted
shock,
\begin{equation}
\bm{\eta}'(0)=\bm{0}.
\label{eq:zero-initial-shock}
\end{equation}
The shock is nevertheless free to move for $t>0$ through its coupling
with the post-shock perturbation dynamics. With the linear evolution
$\bm{q}'_e(t)=\exp(\bm{L}t)\bm{q}'_e(0)$, the gain can be written as
the generalized Rayleigh quotient
\begin{equation}
G(t) =
\frac{\bm{q}'_e(0)^\dagger \exp(\bm{L}t)^\dagger
\bm{P}^\dagger \bm{Q}^\dagger \bm{M}\bm{Q}\bm{P}
\exp(\bm{L}t)\bm{q}'_e(0)}
{\bm{q}'_e(0)^\dagger
\bm{P}^\dagger \bm{Q}^\dagger \bm{M}\bm{Q}\bm{P}
\bm{q}'_e(0)},
\label{eq:gain}
\end{equation}
where $\dagger$ denotes the Hermitian transpose, $\bm{P}$ extracts the
flow perturbation components from $\bm{q}'_e$ excluding the
shock-displacement degrees of freedom, $\bm{Q}$ transforms
conservative variables to Chu variables and $\bm{M}$ is the diagonal
matrix containing the norm weights and finite-volume metric.
Maximization over admissible initial conditions satisfying
$\bm{\eta}'(0)=\bm{0}$ yields a generalized eigenvalue problem that is
solved with a Lanczos iteration. At each time $t$, this maximization
gives the optimal gain
\begin{equation}
G^{\mathrm{opt}}(t)
=
\max_{\substack{\bm{q}'_e(0)\neq\bm{0}\\
                 \bm{\eta}'(0)=\bm{0}}}
G(t).
\label{eq:gopt}
\end{equation}
Repeating the procedure over a range of times identifies the optimal
amplification time and gain
\begin{equation}
t^{\mathrm{opt}}
=
\arg\max_{t>0}
G^{\mathrm{opt}}(t),
\qquad
G^{\mathrm{opt}}_{\max}
=
G^{\mathrm{opt}}(t^{\mathrm{opt}})
=
\max_{t>0}
G^{\mathrm{opt}}(t).
\label{eq:topt}
\end{equation}

\section{Results}
\label{sec:results}

\begin{figure}[tbp]
  \centering
  \subfloat[\label{fig:base-flow-vort}]{%
\begin{minipage}[t]{0.32\textwidth}
\centering
\includegraphics[width=\linewidth]{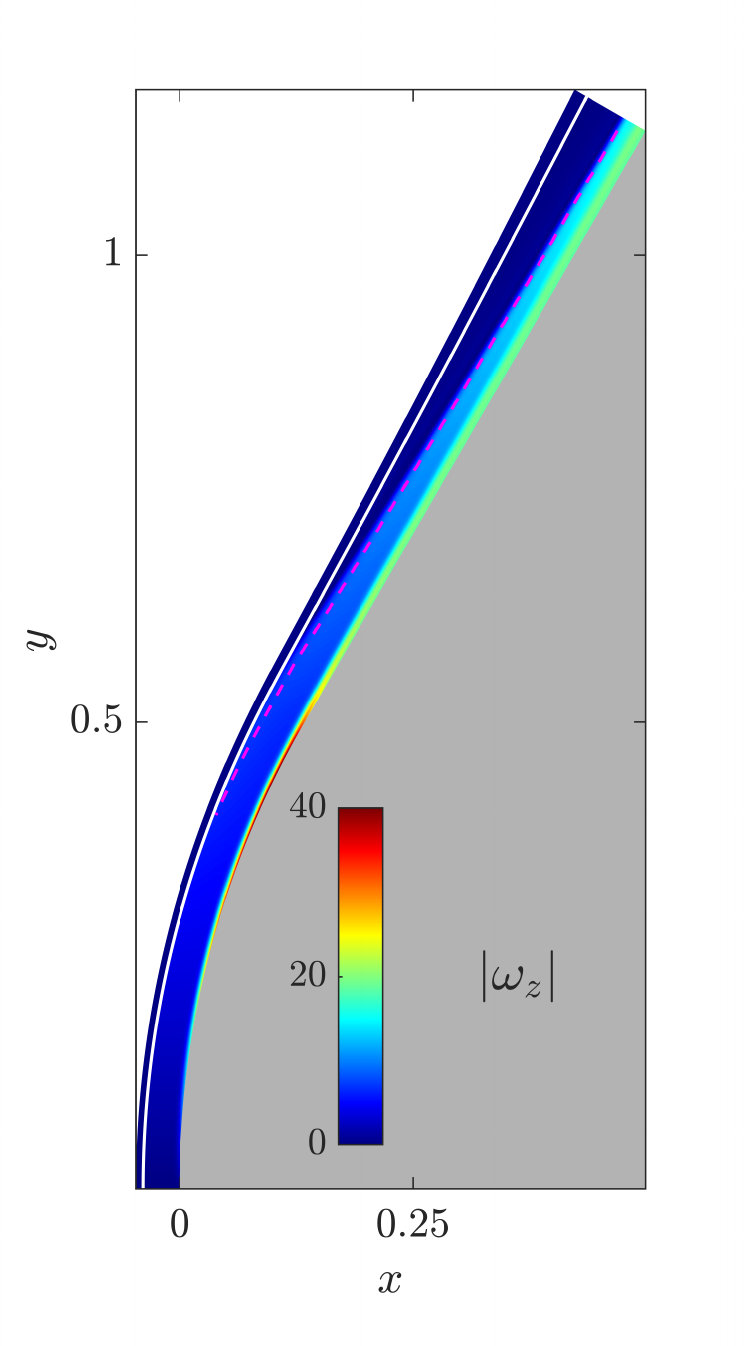}%
\end{minipage}%
}
  \subfloat[\label{fig:base-flow-entropy}]{%
\begin{minipage}[t]{0.32\textwidth}
\centering
\includegraphics[width=\linewidth]{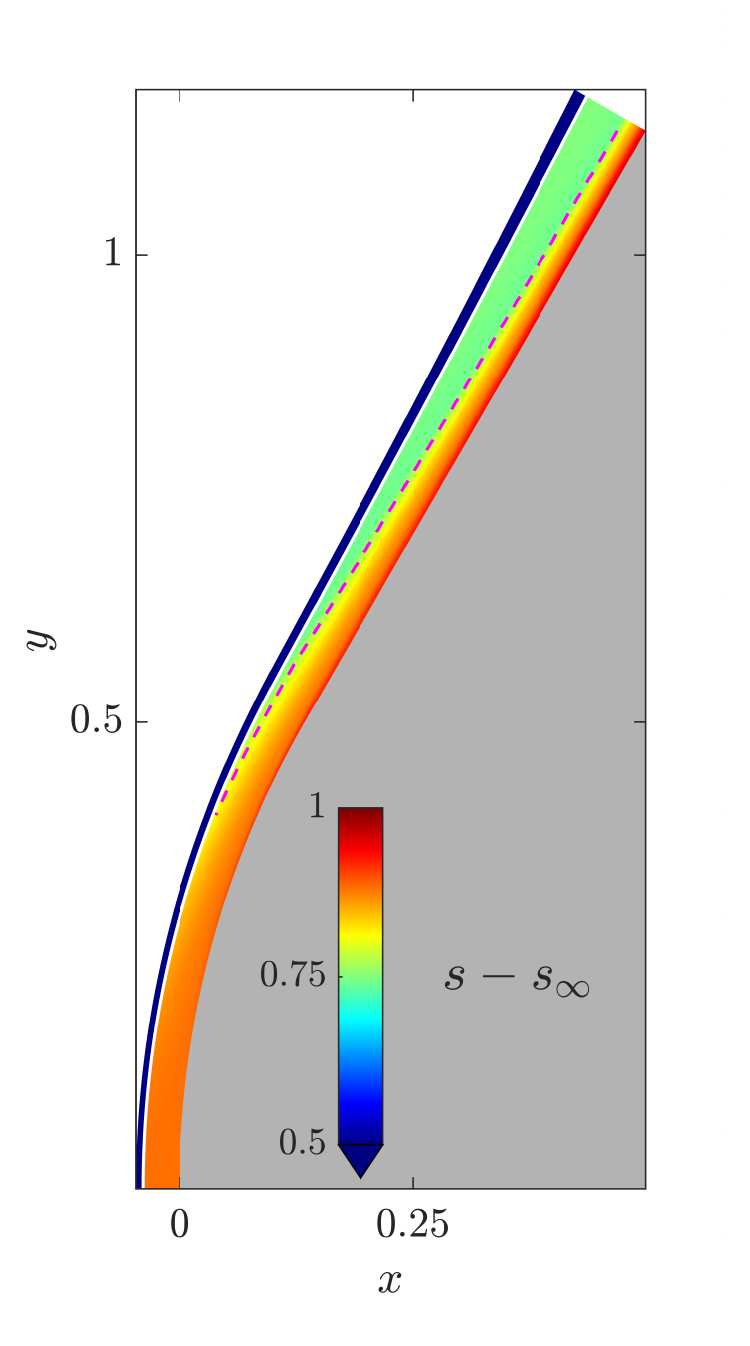}%
\end{minipage}%
}
  \subfloat[\label{fig:base-flow-mach}]{%
\begin{minipage}[t]{0.32\textwidth}
\centering
\includegraphics[width=\linewidth]{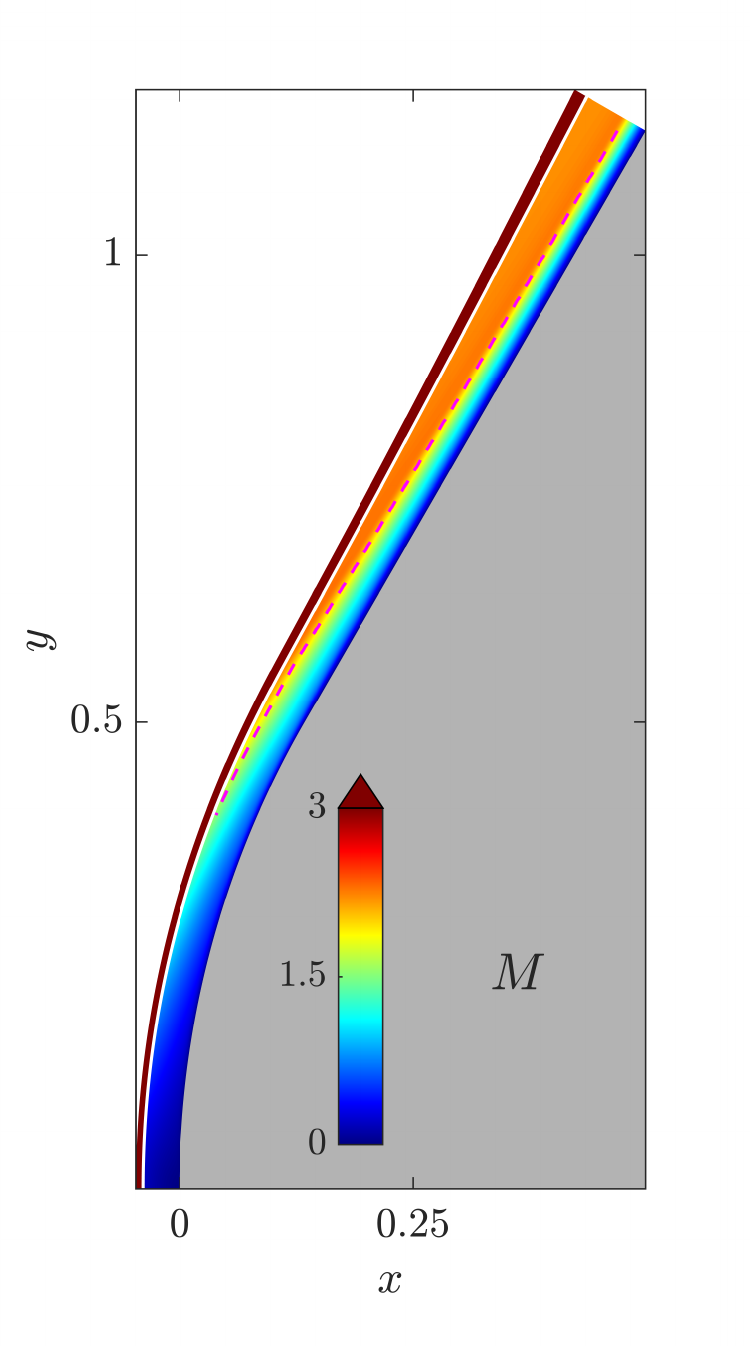}%
\end{minipage}%
}
  \caption{(a) The absolute value of the $z$-component of
    vorticity, (b) entropy relative to the freestream, and
    (c) Mach number for the base flow at the freestream
    conditions listed in Table~\ref{tab:freestream}, with $Re_\infty =
    100\,000$ and $M_\infty = 26.1$.  The capsule is colored in
    gray. The bow shock computed with HYMOR is indicated by a solid
    white line. The shear--entropy layer edge is indicated with a
    dashed magenta line, which marks the location where the entropy
    reaches the uniform value of the flow downstream of the
    approximately straight portion of the oblique shock.  Variables
    are non-dimensionalized as indicated in Sec.~\ref{sec:ns}, while
    entropy is nondimensionalized with freestream entropy
    $s_\infty$. }
  \label{fig:base-flow}
\end{figure}

We present the linear-stability results for the configuration
introduced in Sec.~\ref{sec:problem-setup}. Unless noted otherwise,
the reference case is defined by the freestream conditions listed in
Table~\ref{tab:freestream} with $Re_\infty = 100\,000$. The
corresponding base flow is shown in Fig.~\ref{fig:base-flow}. A
distinguishing feature of this high-enthalpy solution is the small
shock standoff distance, which results from the large density increase
across the bow shock. Moreover, the strong shock curvature produces a
pronounced entropy gradient within the shock layer. Together with the
associated mean shear, this shear--entropy layer gives rise to the
dominant amplification mechanisms discussed below.

A notable property of this base flow is its low post-shock Mach
number. The flow is subsonic over most of the shock layer, and the
Mach number at the edge of the wall boundary layer is $M_e \approx
1.5$ [Fig.~\ref{fig:base-flow-mach}]. This is well below the
range in which Mack's second mode becomes
dominant~\cite{mack1984boundary}, so the near-wall region cannot
support the acoustic instability that has received the most
attention in the hypersonic-transition literature. The resulting
combination of a high-enthalpy entropy--shear layer with a
comparatively low-Mach-number boundary layer has received little
attention to date.

The results are organized into three
parts. Sections~\ref{sec:modal-results}
and~\ref{sec:transient-results} focus on the reference case, in which
the Reynolds number is held fixed to isolate the dominant
mechanisms. Section~\ref{sec:reynolds-dependence} then examines the
effect of varying $Re_\infty$ to determine how these mechanisms scale
with Reynolds number.

\subsection{Modal stability analysis}
\label{sec:modal-results}

The global modal analysis indicates that the reference flow is
temporally stable. The leading eigenvalues are listed in
Table~\ref{tab:eigenvalues}.  The least stable mode has
$\lambda=-0.300$, so the perturbation evolves as
$\bm{q}'_e(t)=\exp(\lambda t)\bm{q}'_e(0)$. This behavior is consistent
with previous studies of blunt or blunted hypersonic forebodies, which
found that the most relevant instabilities in these flows are
typically convective rather than temporally
global~\cite{paredes2017blunt,paredes2019nose,paredes2020mechanism}.

\begin{figure}[tbp]
\centering
\subfloat[\label{fig:modal-1}]{%
\begin{minipage}[t]{0.32\textwidth}
\centering
\includegraphics[width=\linewidth]{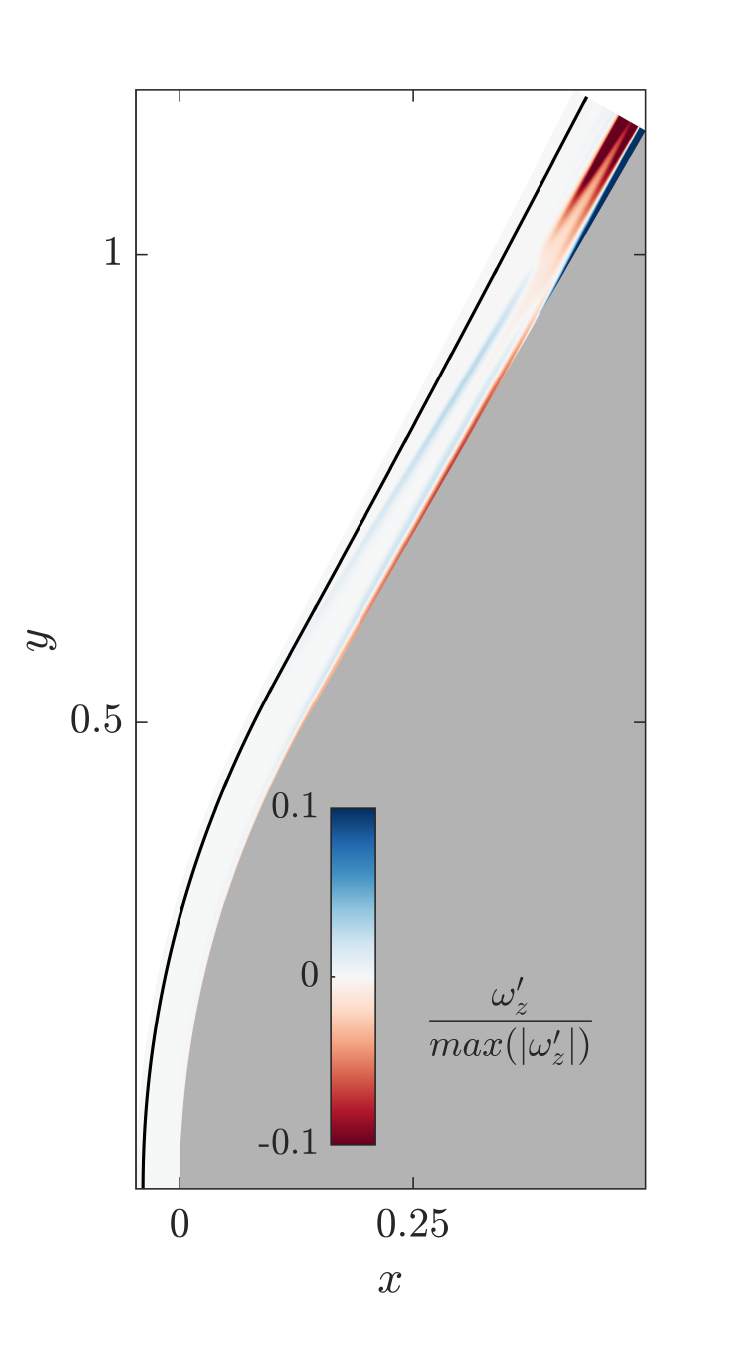}%
\end{minipage}%
}
\subfloat[\label{fig:modal-2}]{%
\begin{minipage}[t]{0.32\textwidth}
\centering
\includegraphics[width=\linewidth]{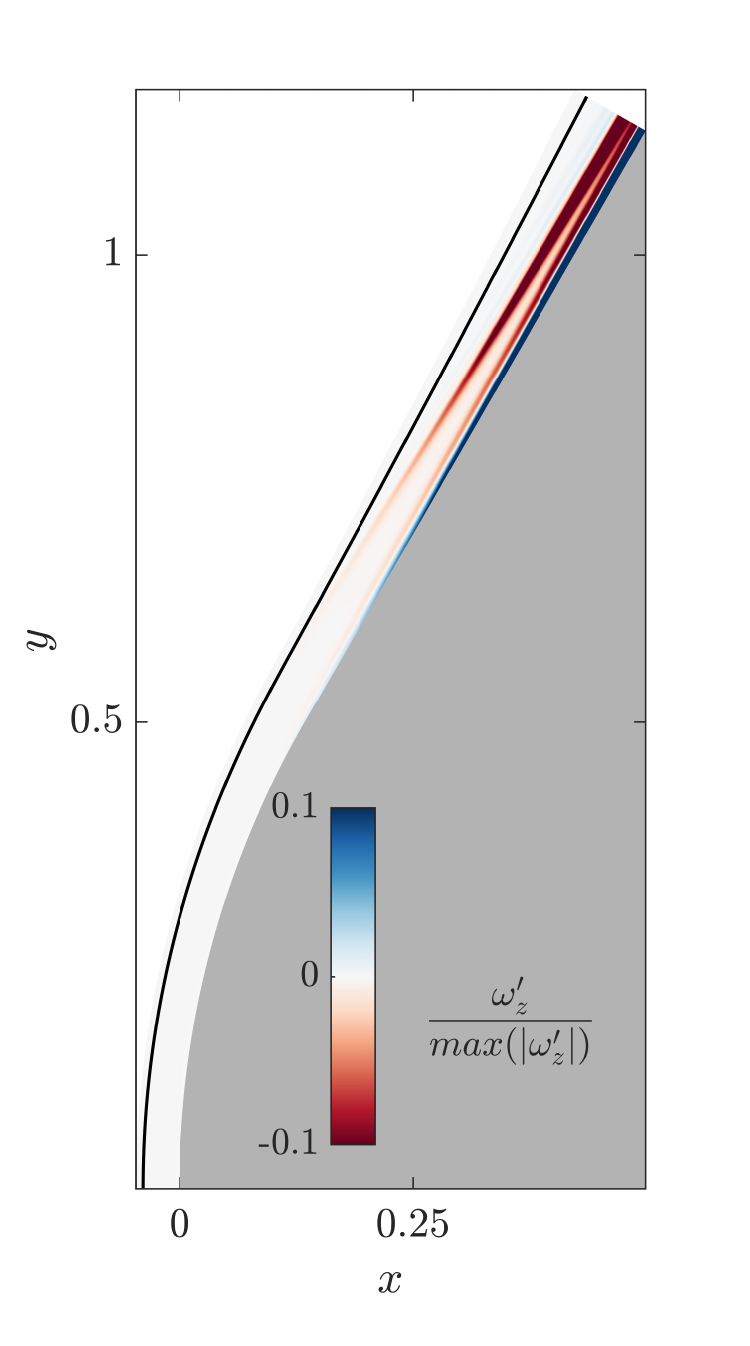}%
\end{minipage}%
}
\subfloat[\label{fig:modal-3}]{%
\begin{minipage}[t]{0.32\textwidth}
\centering
\includegraphics[width=\linewidth]{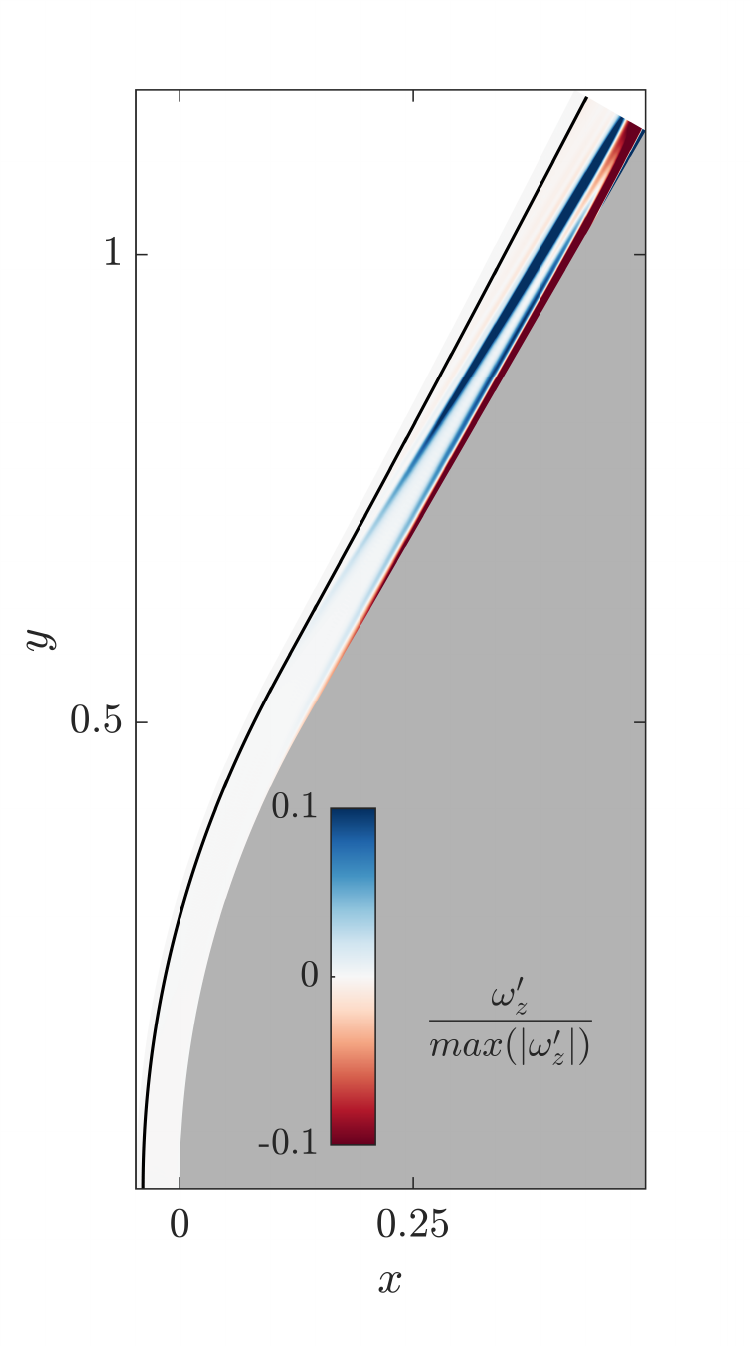}%
\end{minipage}%
}
\caption{Disturbance-vorticity fields associated with the three least
  stable global modes of the base flow defined by the freestream
  conditions in Table~\ref{tab:freestream}, at $Re_\infty = 100\,000$
  and $M_\infty = 26.1$. The bow shock computed with HYMOR is
  indicated by a solid black line, a convention that will be
  maintained throughout the rest of the paper.  (a) Mode
  1. (b) Mode 2. (c) Mode 3.}
\label{fig:modal-modes}
\end{figure}

The spatial structure of the three least stable eigenmodes is shown in
Fig.~\ref{fig:modal-modes} through their disturbance vorticity. The
modes are concentrated within the two shear regions visible in the
base flow of Fig.~\ref{fig:base-flow-vort}: the shear--entropy
layer edge generated by the curved bow shock and the inner wall boundary
layer. None of these axisymmetric disturbance structures exhibits temporal growth. 
\begin{table}[t]
\caption{Real and imaginary parts of the leading eigenvalues from the
    global modal analysis for the conditions in
    Table~\ref{tab:freestream}, with $Re_\infty = 100\,000$
    and $M_\infty = 26.1$.}
\label{tab:eigenvalues}
\centering
\begin{ruledtabular}
\begin{tabular}{lcccc}
    Mode & 1 & 2 & 3 & 4 \\
    \colrule
    $\operatorname{Re}(\lambda)$
    & $-0.300$ & $-0.345$ & $-0.487$ & $-0.487$ \\
    $\operatorname{Im}(\lambda)$
    & $0.000$ & $0.000$ & $0.068$ & $-0.068$ \\
\end{tabular}
\end{ruledtabular}
\end{table}

\subsection{Transient growth}
\label{sec:transient-results}

\begin{figure}[tbp]
\centering
\includegraphics[width=0.5\textwidth]{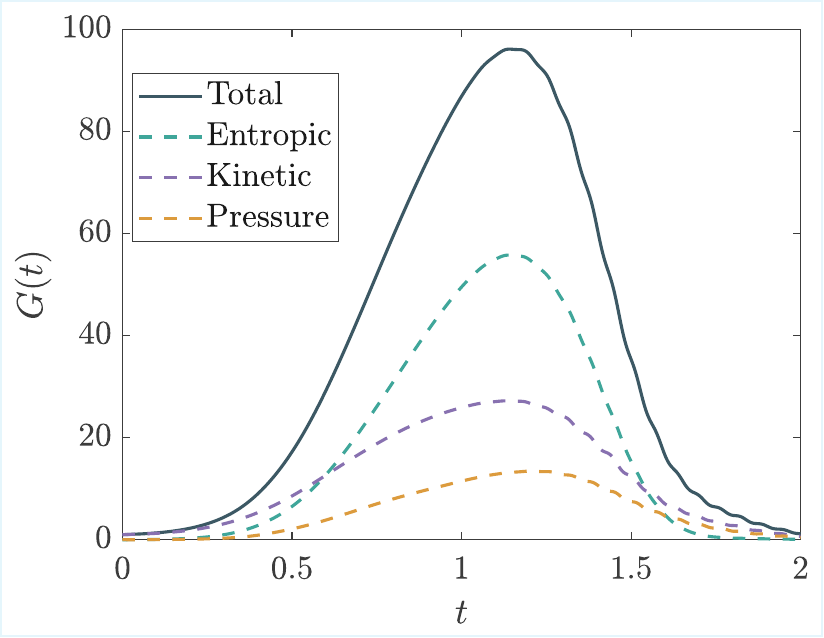}
\caption{Temporal evolution of gain associated with Chu's energy
  components for the optimal initial post-shock disturbance. The
  maximum gain is $G_{\max}^\text{opt}=96.4$. The base flow
  corresponds to Table~\ref{tab:freestream} at $Re_\infty = 100\,000$
  and $M_\infty = 26.1$.}
\label{fig:energy-growth}
\end{figure}

The non-modal analysis reveals a different picture. Although the
axisymmetric eigenspectrum is stable, the optimal axisymmetric
disturbance achieves an energy gain close to $100$, as shown in
Fig.~\ref{fig:energy-growth}. The amplification is dominated by the
kinetic and entropic contributions, while the pressure component
remains secondary.  Gains of comparable order have been reported in
other hypersonic transient-growth studies, such as blunt-cone and
elliptic-cone configurations (see, e.g., Refs.~\cite{paredes2020mechanism,
  quintanilha2022transient}), indicating that transient growth can
provide a robust linear-amplification pathway across a wide range of
high-speed flows.

\begin{figure}[tbp]
\centering
\subfloat[\label{fig:optimal-initial-wide}]{%
\begin{minipage}[t]{0.32\textwidth}
\centering
\includegraphics[width=\linewidth]{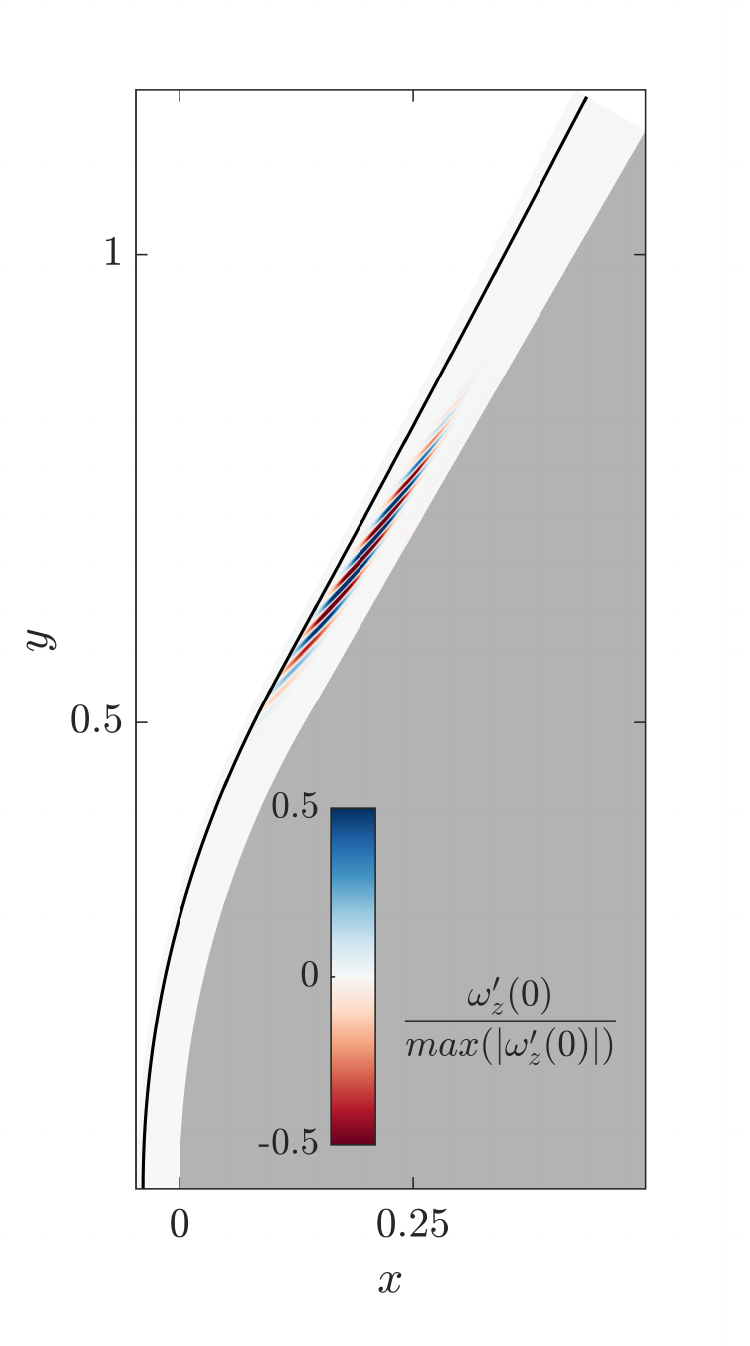}%
\end{minipage}%
}
\subfloat[\label{fig:optimal-initial-zoom}]{%
\begin{minipage}[t]{0.32\textwidth}
\centering
\includegraphics[width=\linewidth]{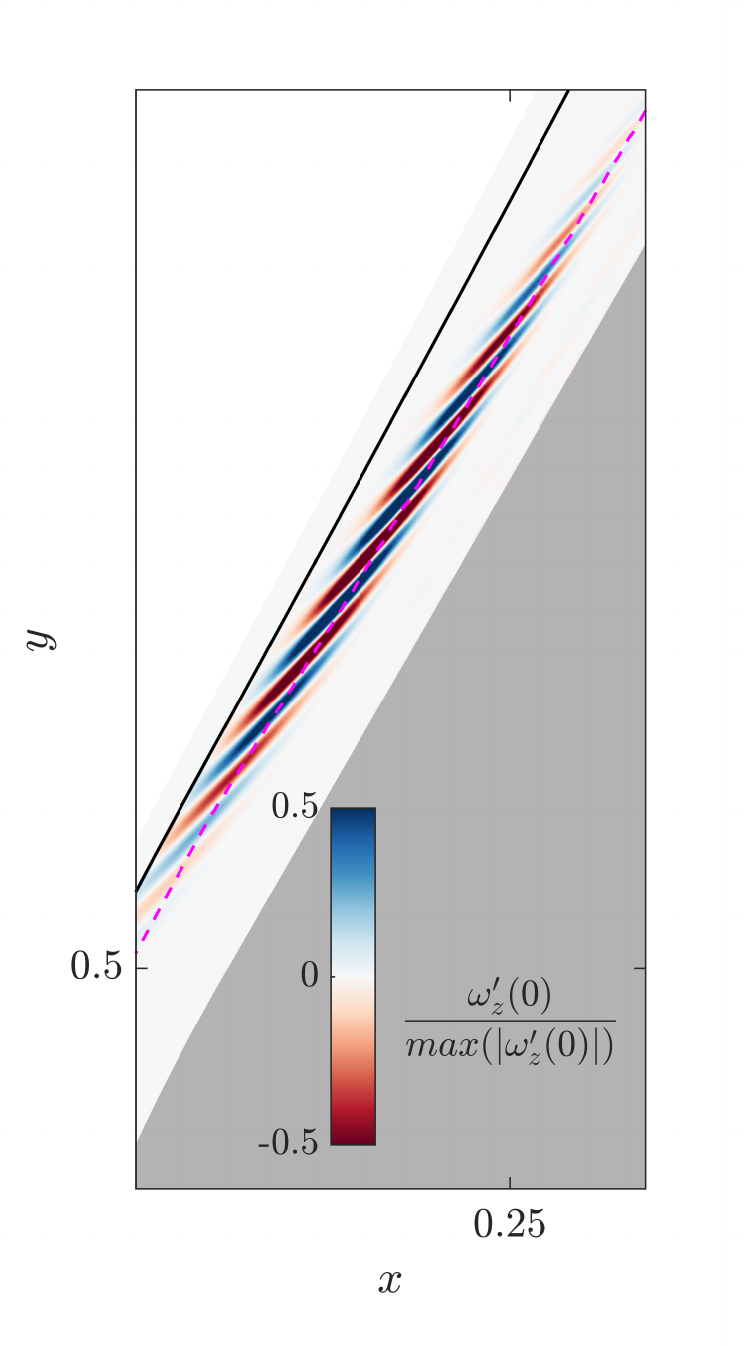}%
\end{minipage}%
}
\caption{Optimal disturbance at the initial time for the reference
  case. The maximum gain is $G_{\max}^\text{opt}=96.4$. Base flow
  computed from Table~\ref{tab:freestream} at $Re_\infty = 100\,000$
  and $M_\infty = 26.1$. (a) Disturbance
  vorticity. (b) Zoom of the disturbance structure. The
  shear--entropy layer edge from the base flow
  (Fig.~\ref{fig:base-flow}) is indicated with a dashed magenta line.}
\label{fig:optimal-initial}
\end{figure}

\begin{figure}[tbp]
\centering
\subfloat[\label{fig:optimal-topt-wide}]{%
\begin{minipage}[t]{0.32\textwidth}
\centering
\includegraphics[width=\linewidth]{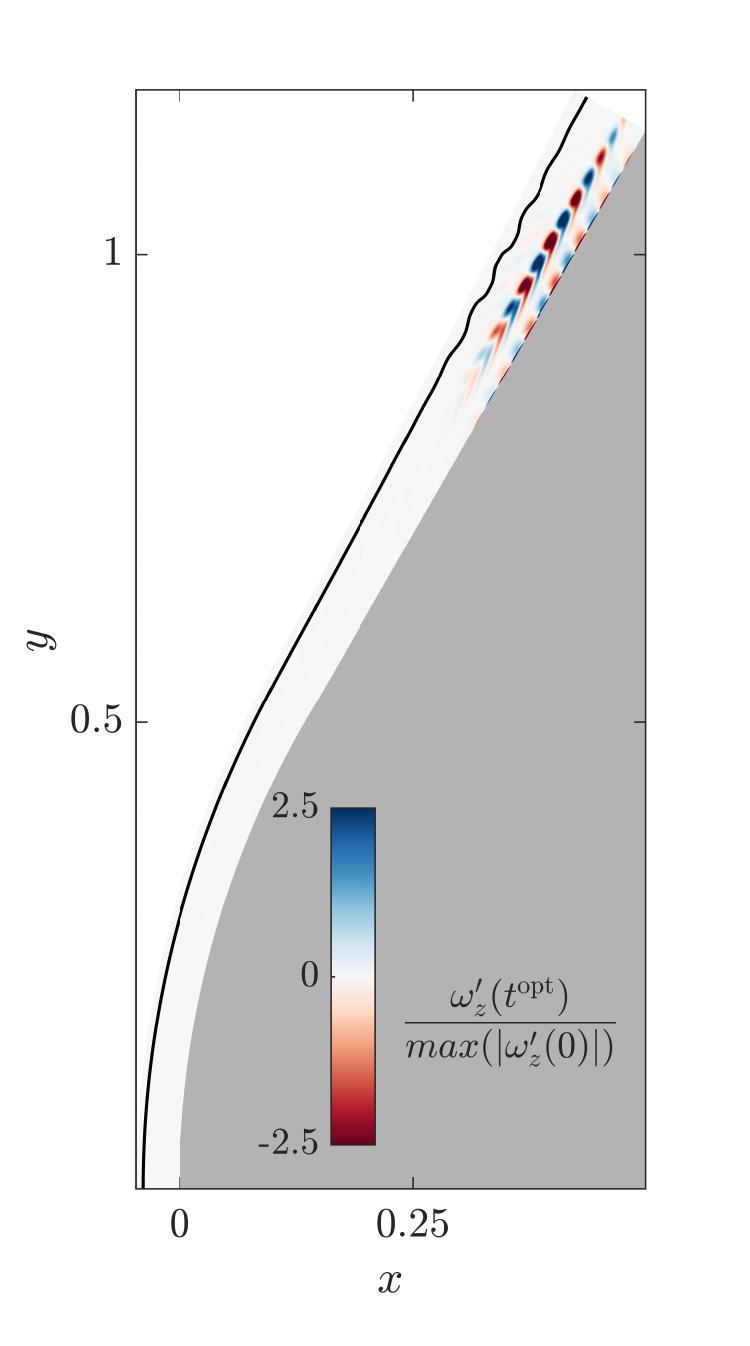}%
\end{minipage}%
}
\subfloat[\label{fig:optimal-topt-vort}]{%
\begin{minipage}[t]{0.32\textwidth}
\centering
\includegraphics[width=\linewidth]{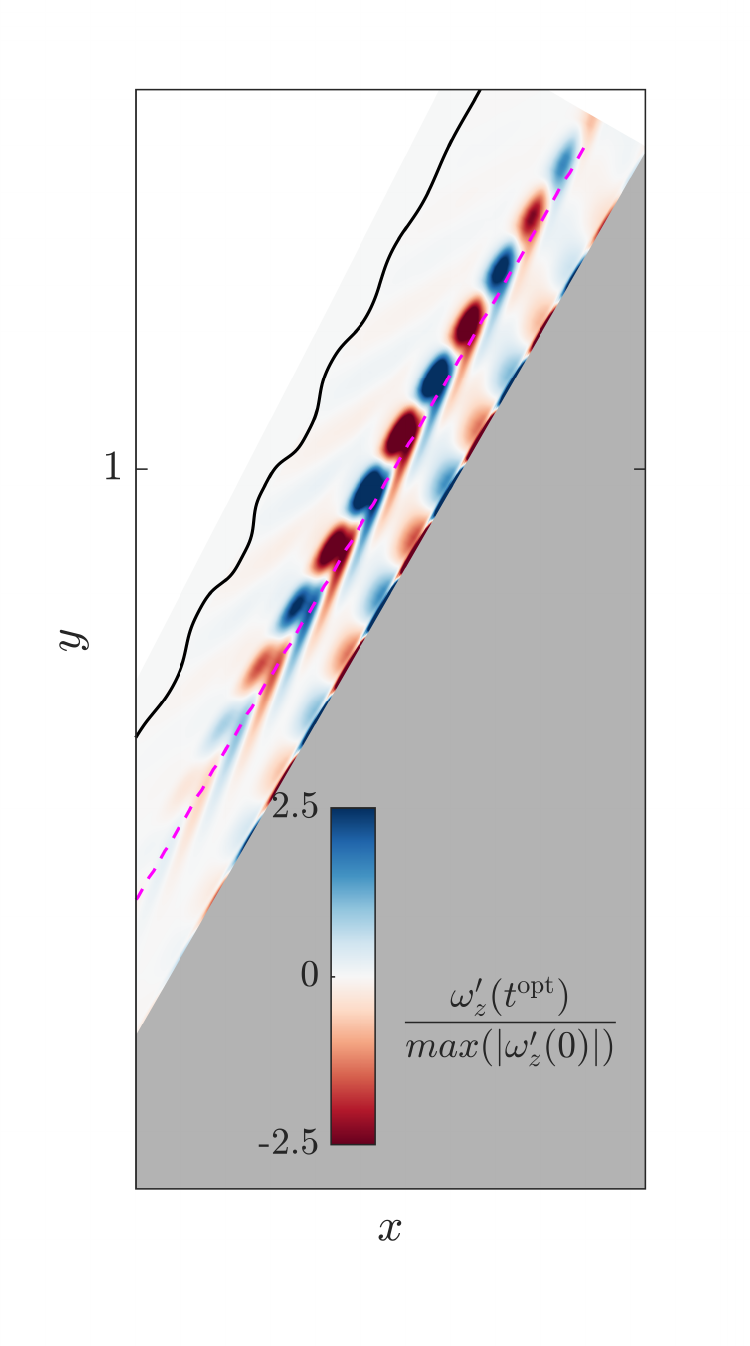}%
\end{minipage}%
}
\subfloat[\label{fig:optimal-entropy}]{%
\begin{minipage}[t]{0.32\textwidth}
\centering
\includegraphics[width=\linewidth]{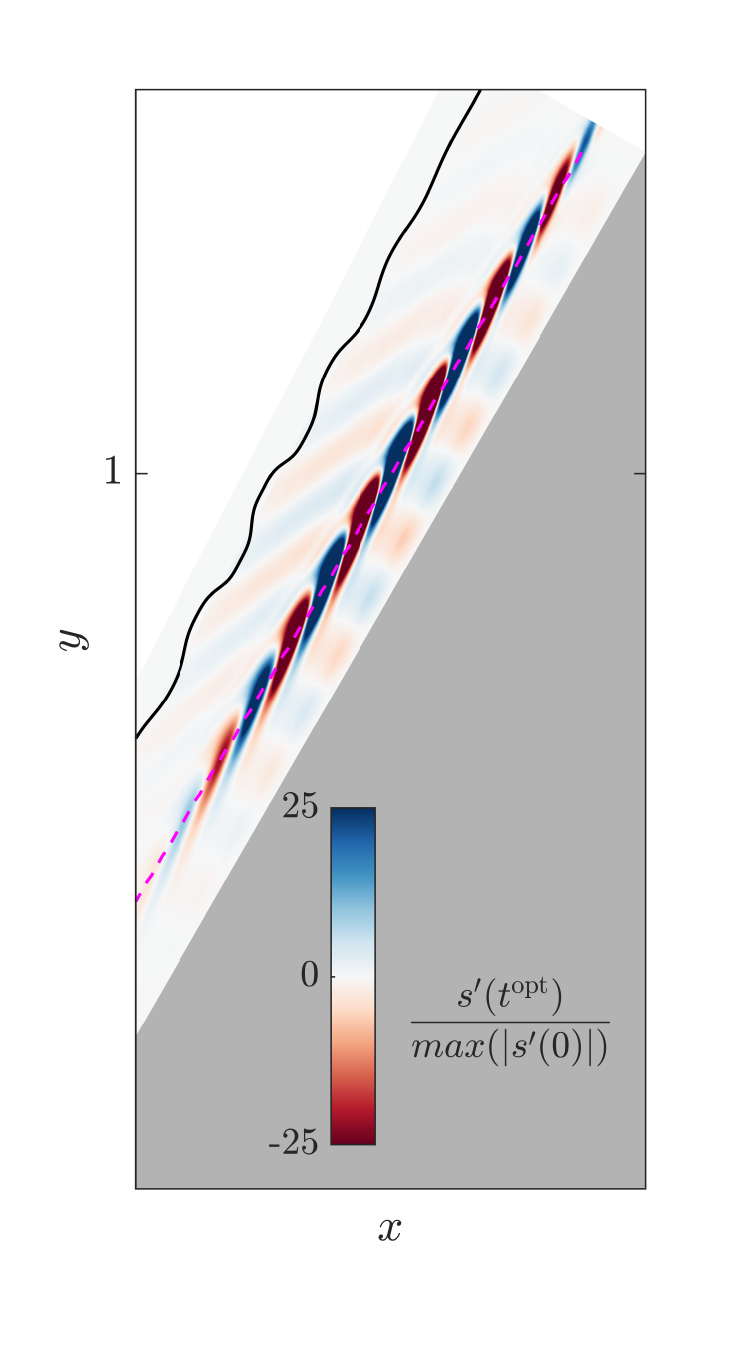}%
\end{minipage}%
}
\caption{Optimal disturbance at the time of maximum amplification $t^\text{opt}$ for
  the reference case. The maximum gain is $G_{\max}^\text{opt}=96.4$. Base
  flow computed from Table~\ref{tab:freestream} at $Re_\infty =
  100\,000$ and $M_\infty = 26.1$. The shear--entropy layer edge from 
  the base flow (Fig.~\ref{fig:base-flow}) is indicated 
  with a dashed magenta line. (a) Disturbance
  vorticity.  (b) Zoom of the disturbance-vorticity
  field. (c) Zoom of the disturbance-entropy field.}
\label{fig:optimal-topt}
\end{figure}

The structure of the optimal disturbance is shown at the initial time
and at the time of maximum amplification in
Figs.~\ref{fig:optimal-initial} and~\ref{fig:optimal-topt},
respectively. At $t=0$, the perturbation is concentrated at the outer edge of the
shear--entropy layer generated by the curved shock. The initial wave
packets are tilted against the mean shear, so that downstream
advection progressively reorients them and enables energy extraction
from the base flow. This evolution is reminiscent of the Orr
mechanism, but here the disturbance continues to draw energy while it
is convected along the entropy layer.

As the perturbation amplifies, it remains largely confined to the
entropy layer and begins to induce visible oscillations of the bow
shock (Fig.~\ref{fig:optimal-topt}). This evolution also generates
large entropy fluctuations through interaction with the strong mean
entropy gradient, as shown in
Fig.~\ref{fig:optimal-entropy}. Because the entropy layer narrows in
the downstream direction, the amplified wave packet is progressively
displaced toward the wall. This behavior suggests a possible route to
transition in which disturbances initially grow within the entropy
layer and subsequently interact with the wall boundary layer once they
reach finite amplitude.

To clarify the amplification process, we examine the kinetic and
entropic energy budgets associated with the Chu norm
integrated over the post-shock region $V_D$. We denote $\left\langle f
\right\rangle_D \equiv \int_{V_D} f\,dV$, and define the kinetic
energy contribution to Chu's norm as $E^k = \left\langle 1/2\rho_0
u'_i u'_i \right\rangle_D$. The corresponding kinetic-energy budget is
\begin{equation}
  \frac{dE^k}{dt} 
  =
  \left\langle \mathcal{A}^k \right\rangle_D
  +
  \left\langle \mathcal{P}^k \right\rangle_D
  +
  \left\langle \Pi_d^k \right\rangle_D
  +
  \left\langle \mathcal{T}^k \right\rangle_D
  +
  \left\langle \mathcal{D}^k \right\rangle_D,
  \label{eq:kinetic-budget-main}
\end{equation}
where $\mathcal{A}^k$ is advection of perturbation kinetic energy by
the base flow, $\mathcal{P}^k$ is production by base-flow velocity
gradients, $\Pi_d^k$ is pressure--dilatation, $\mathcal{T}^k$ is the
combined pressure and viscous transport, and $\mathcal{D}^k$ is
viscous dissipation.  The entropic budget for $E^s= \left\langle
(\gamma_0^*-1)p_0/(2\gamma_0^*) (s'/R_{g,0})^2 \right\rangle_D$ is
\begin{equation}
  \frac{dE^s}{dt} 
  =  
  \left\langle \mathcal{A}^s \right\rangle_D
  +
  \left\langle \mathcal{P}^s \right\rangle_D
  +
  \left\langle \mathcal{T}^s \right\rangle_D
  +
  \left\langle \mathcal{D}^s \right\rangle_D
  +
  \left\langle \mathcal{S}^s \right\rangle_D ,
  \label{eq:entropic-budget-main}
\end{equation}
where $\mathcal{A}^s$ is base-flow advection of entropic energy,
$\mathcal{P}^s$ is production by perturbation advection of the
base-flow entropy gradient, $\mathcal{T}^s$ is heat-flux transport,
$\mathcal{D}^s$ is diffusive dissipation, and $\mathcal{S}^s$ is the
source associated with viscous heating.  For compactness, the brackets
$\langle\cdot\rangle_D$ are omitted below when referring to
domain-integrated budget terms.  The definitions of the budget terms
are provided in Appendix~\ref{app:budget}.

The kinetic budget in
Fig.~\ref{fig:kinetic-budget} shows that the dominant positive
contribution is the production term $\mathcal{P}^k$, primarily through
the Reynolds-stress component
\begin{equation*}
\mathcal{P}_u^k = -\rho_0 u_i'u_j'\frac{\partial u_{i0}}{\partial
  x_j}.
\end{equation*}
This is consistent with the structure of the optimal initial condition
in Fig.~\ref{fig:optimal-initial}: the disturbance is concentrated
in the high-shear portion of the shear--entropy layer created by
bow-shock curvature.  This is consistent with the shock-layer
structure highlighted experimentally by \citet{hornung2001shock}. At
$Re_\infty = 100\,000$, the viscous-dissipation term $\mathcal{D}^k$
is comparatively small and cannot offset the production. The main
processes that limit the net gain are advection and transport, which
become strongly negative once the wave packet exits the downstream
boundary.

The entropic budget in Fig.~\ref{fig:entropic-budget} displays an
analogous sequence. The dominant source is the production term
$\mathcal{P}^s$, especially the contribution associated with velocity
fluctuations,
\begin{equation*}
\mathcal{P}_u^s = -\frac{(\gamma_0^*-1)p_0}{\gamma_0^*R_{g,0}^2}
u_j's'\frac{\partial s_0}{\partial x_j}.
\end{equation*}
The entropic energy peaks later than the kinetic energy, consistent
with the two-stage amplification process indicated by the production
terms.  Hence, the velocity fluctuations are first amplified by
mean-shear production and then generate entropy fluctuations as they
are advected across the background entropy gradient. As in the kinetic
budget, advection and transport dominate the subsequent decay because
the disturbance leaves the computational domain. This behavior
suggests that amplification is predominantly convective rather than
driven by a temporally unstable global mode.
\begin{figure}[tbp]
\centering
\subfloat[\label{fig:kinetic-budget}]{%
\begin{minipage}[t]{0.49\textwidth}
\centering
\includegraphics[width=\linewidth]{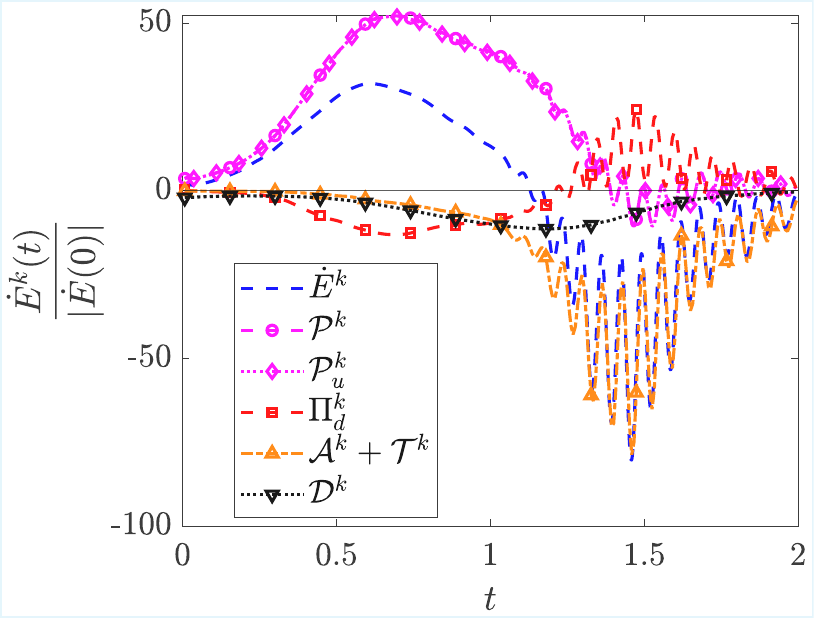}%
\end{minipage}%
}
\subfloat[\label{fig:entropic-budget}]{%
\begin{minipage}[t]{0.49\textwidth}
\centering
\includegraphics[width=\linewidth]{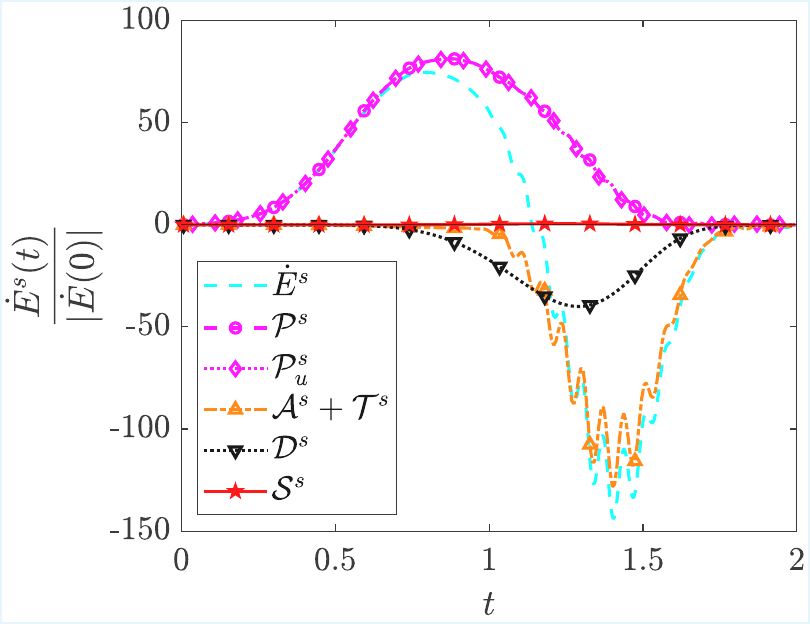}%
\end{minipage}%
}
\caption{Energy budgets for the optimal post-shock disturbance of the
  reference case, for which $G_{\max}^\text{opt}=96.4$. The base flow
  corresponds to Table~\ref{tab:freestream} at $Re_\infty = 100\,000$
  and $M_\infty = 26.1$.  (a) Kinetic energy
  budget. (b) Entropic energy budget. The plotted curves are
  the rate of change of energy $\dot{E}$, total production
  $\mathcal{P}$, its velocity-induced component $\mathcal{P}_u$,
  pressure--dilatation $\Pi_d$, the combined advection-plus-transport
  contribution $\mathcal{A}+\mathcal{T}$, dissipation $\mathcal{D}$,
  and the entropy source $\mathcal{S}$. Superscripts $k$ and $s$
  denote kinetic and entropic contributions, respectively. The high
  frequency oscillations in the results arise from wave packets
  leaving the downstream domain.}
\label{fig:budgets-reference}
\end{figure}

\subsection{Reynolds-number dependence}
\label{sec:reynolds-dependence}

\begin{figure}[tbp]
  \centering
\subfloat[]{%
\includegraphics[width=0.48\linewidth]{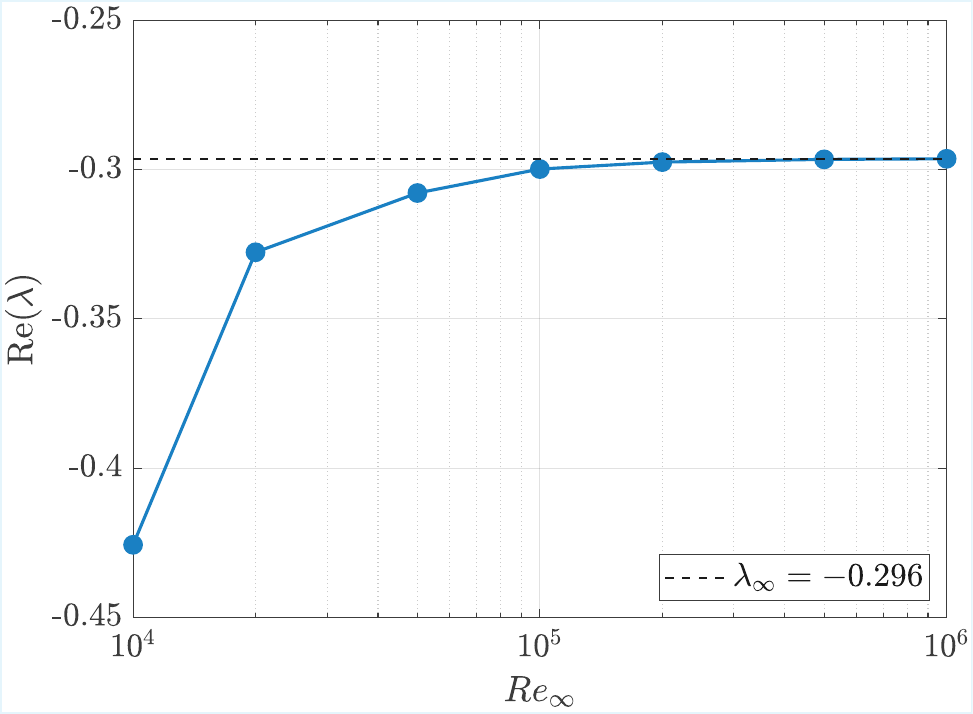}%
}
\subfloat[]{%
\includegraphics[width=0.47\linewidth]{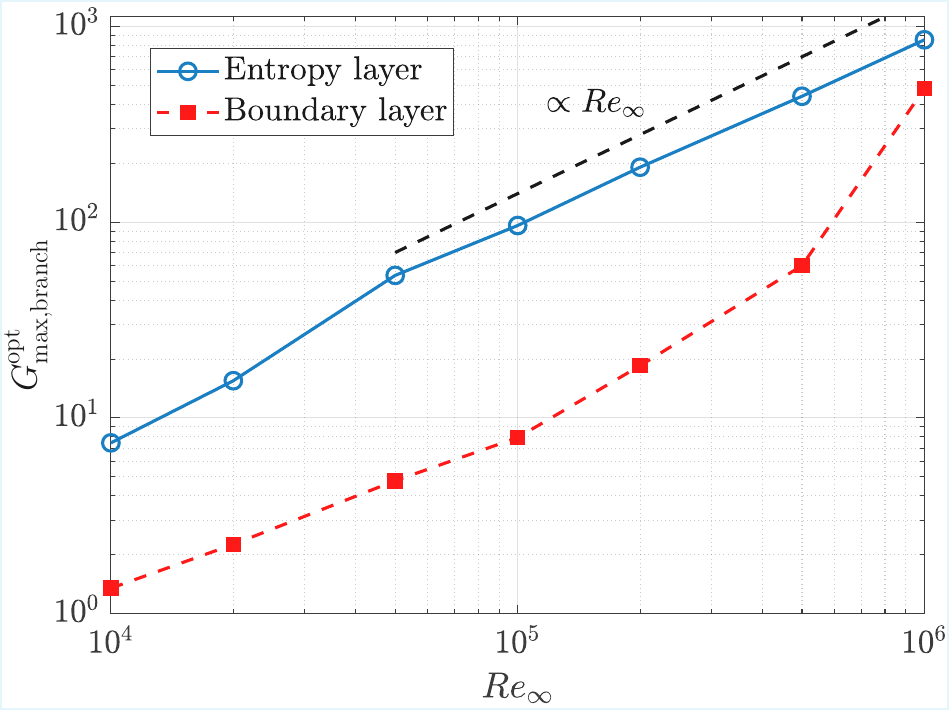}%
}
    \caption{(a) Global stability analysis: real part of the
      least stable eigenvalue as a function of $Re_\infty$.
      (b) Transient growth: branchwise optimal gains as a
      function of $Re_\infty$. Blue circles denote the entropy-layer
      branch, $G_{\max,\mathrm{EL}}^\text{opt}$, and red squares
      denote the adiabatic boundary-layer-localized branch,
      $G_{\max,\mathrm{BL}}^\text{opt}$. The dashed line indicates the
      reference scaling $G_{\max,\mathrm{EL}}^\text{opt}\propto
      Re_\infty$. The base flow is computed from
      Table~\ref{tab:freestream} at $M_\infty = 26.1$.}
    \label{fig:modal-gain-vs-re}
\end{figure}

We examine how the modal behavior varies with Reynolds number. For all
values considered, the flow remains globally stable.
Figure~\ref{fig:modal-gain-vs-re}(a) plots the real part of the least
stable eigenvalue as a function of $Re_\infty$. The eigenvalue stays
in the stable half-plane and approaches the asymptotic value
$\lambda_\infty\approx-0.296$ as the Reynolds number increases. This
is consistent with the fact that the instability mechanisms usually
reported for these flows are convective in space and therefore do not
appear as temporal global modes.

The non-modal behavior, on the other hand, shows clear dependence on
$Re_\infty$. Figure~\ref{fig:modal-gain-vs-re}(b) shows the branchwise
optimal gains over the Reynolds-number range examined. For
sufficiently large Reynolds numbers, the entropy-layer branch scales
approximately linearly with $Re_\infty$. A second branch, localized in
the wall boundary layer, yields gains comparable to those of the
entropy-layer branch at $Re_\infty \approx 10^6$.

\begin{figure}[tbp]
    \centering
    \subfloat[\label{fig:re-mode-20000}]{%
\begin{minipage}[t]{0.32\textwidth}
\centering
\includegraphics[width=\linewidth]{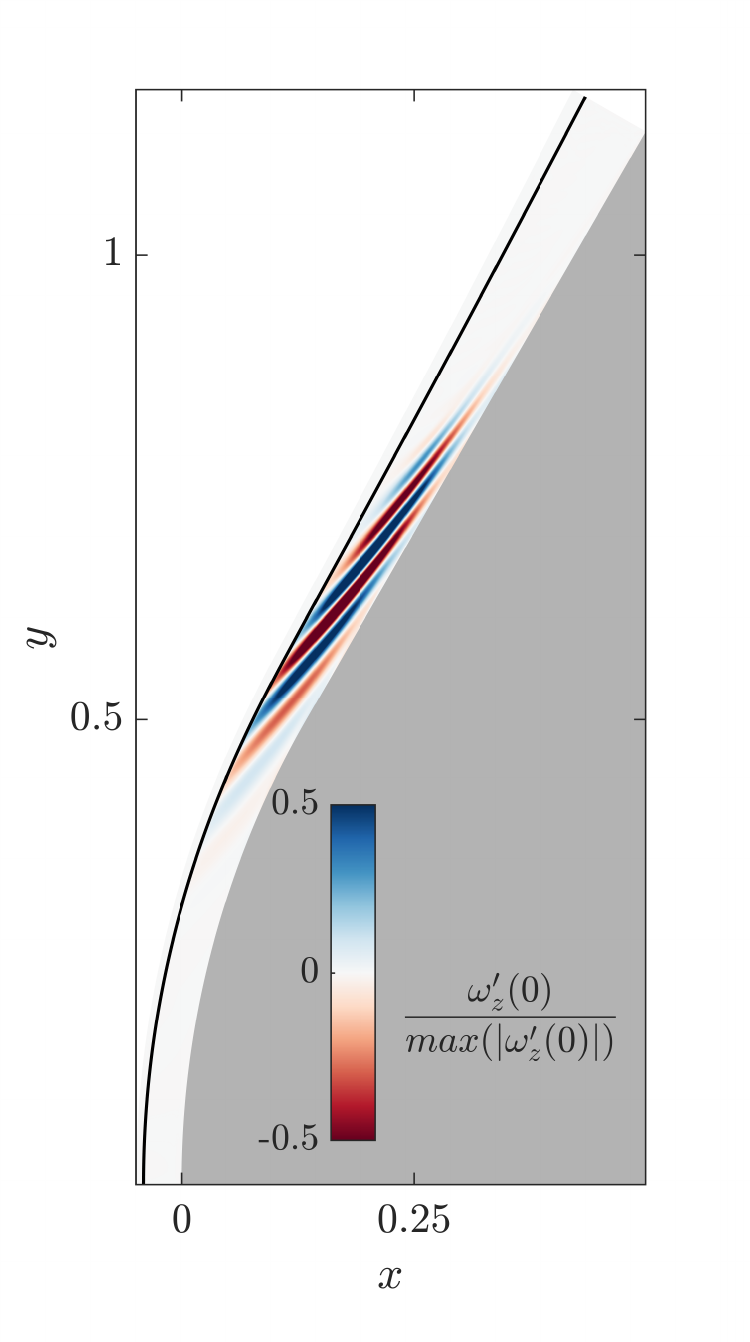}%
\end{minipage}%
}
    \subfloat[\label{fig:re-mode-100000}]{%
\begin{minipage}[t]{0.32\textwidth}
\centering
\includegraphics[width=\linewidth]{images/Re_100000_MSL/Non-Modal_Downstream/non_modal_down_100000_vort_0.pdf}%
\end{minipage}%
}
    \subfloat[\label{fig:re-mode-500000}]{%
\begin{minipage}[t]{0.32\textwidth}
\centering
\includegraphics[width=\linewidth]{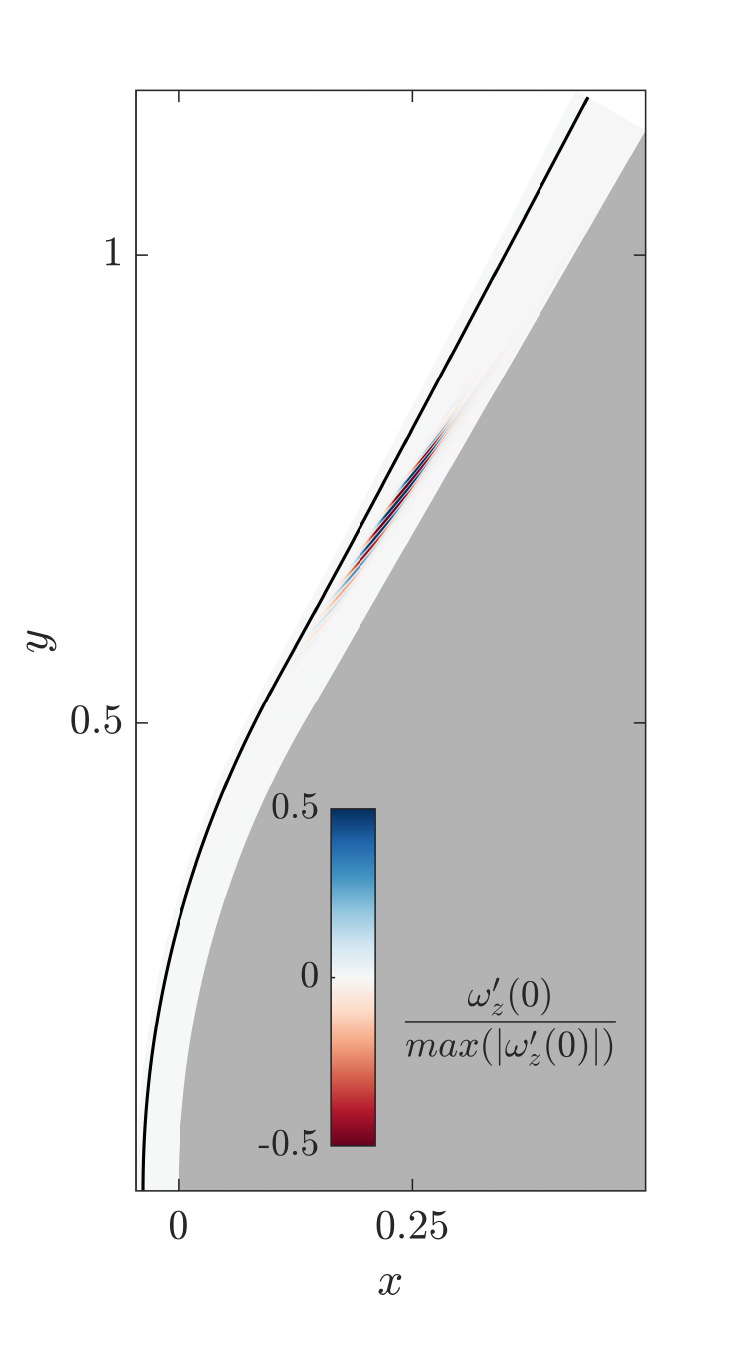}%
\end{minipage}%
}
    \caption{Optimal initial disturbance vorticity for the
      entropy-layer branch at three Reynolds numbers. The base flow is
      computed from Table~\ref{tab:freestream} at $M_\infty =
      26.1$ for (a) $Re_\infty = 20\,000$, (b)
      $Re_\infty = 100\,000$, and (c) $Re_\infty = 500\,000$.}
    \label{fig:initial-re-sweep}
\end{figure}

\begin{figure}[tbp]
  \centering
  \subfloat[]{%
\includegraphics[width=0.32\textwidth]{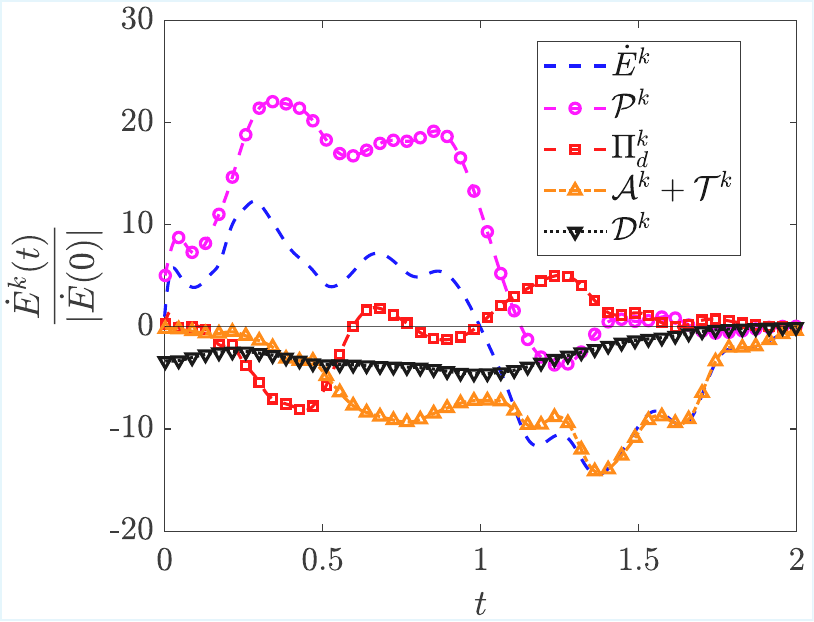}%
}
\subfloat[]{%
\includegraphics[width=0.32\textwidth]{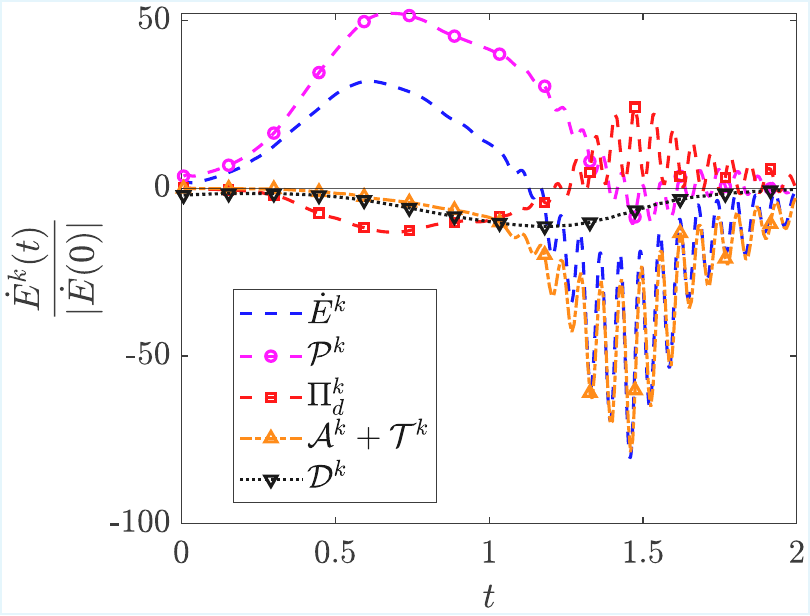}%
}
\subfloat[]{%
\includegraphics[width=0.32\textwidth]{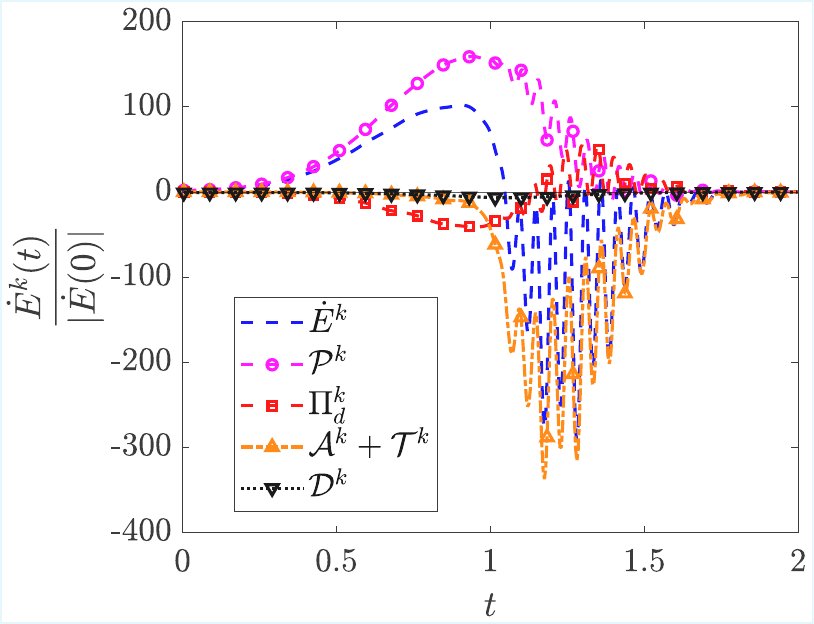}%
}
    \caption{Kinetic-energy budgets of the optimal entropy-layer
      disturbance for three Reynolds numbers. The base flow is
      computed from Table~\ref{tab:freestream} at $M_\infty =
      26.1$. (a) $Re_\infty = 20\,000$. (b)
      $Re_\infty = 100\,000$. (c) $Re_\infty = 500\,000$.}
    \label{fig:re-budgets}
\end{figure}

The physical origin of the scaling of the entropy-layer branch is
visible in Fig.~\ref{fig:initial-re-sweep}, which presents the
optimal initial disturbance at three representative Reynolds numbers.
As $Re_\infty$ increases, the wave packets become progressively finer
and their wavelength decreases, while remaining localized within the
same high-shear region near the outer edge of the shear--entropy
layer. In other words, increasing the Reynolds number reduces the
scale of the optimal structure without altering the region in which
the disturbance is concentrated. The corresponding kinetic-energy
budgets, shown in Fig.~\ref{fig:re-budgets}, confirm that the
mechanism identified in Sec.~\ref{sec:transient-results} remains
unchanged over the range of $Re_\infty$ considered. In all cases, the
dominant positive contribution is the Reynolds-stress production term
$\mathcal{P}_u^k$, whereas the main negative contributions arise when
advection and transport carry the amplified wave packet out of the
computational domain. The magnitude of the production term increases
with $Re_\infty$, consistent with the linear growth of
$G_{\max}^{\mathrm{opt}}$.

%
\begin{figure}[tbp]
    \centering
    \subfloat[\label{fig:bl-branch-wide}]{%
\begin{minipage}[t]{0.32\textwidth}
\centering
\includegraphics[width=\linewidth]{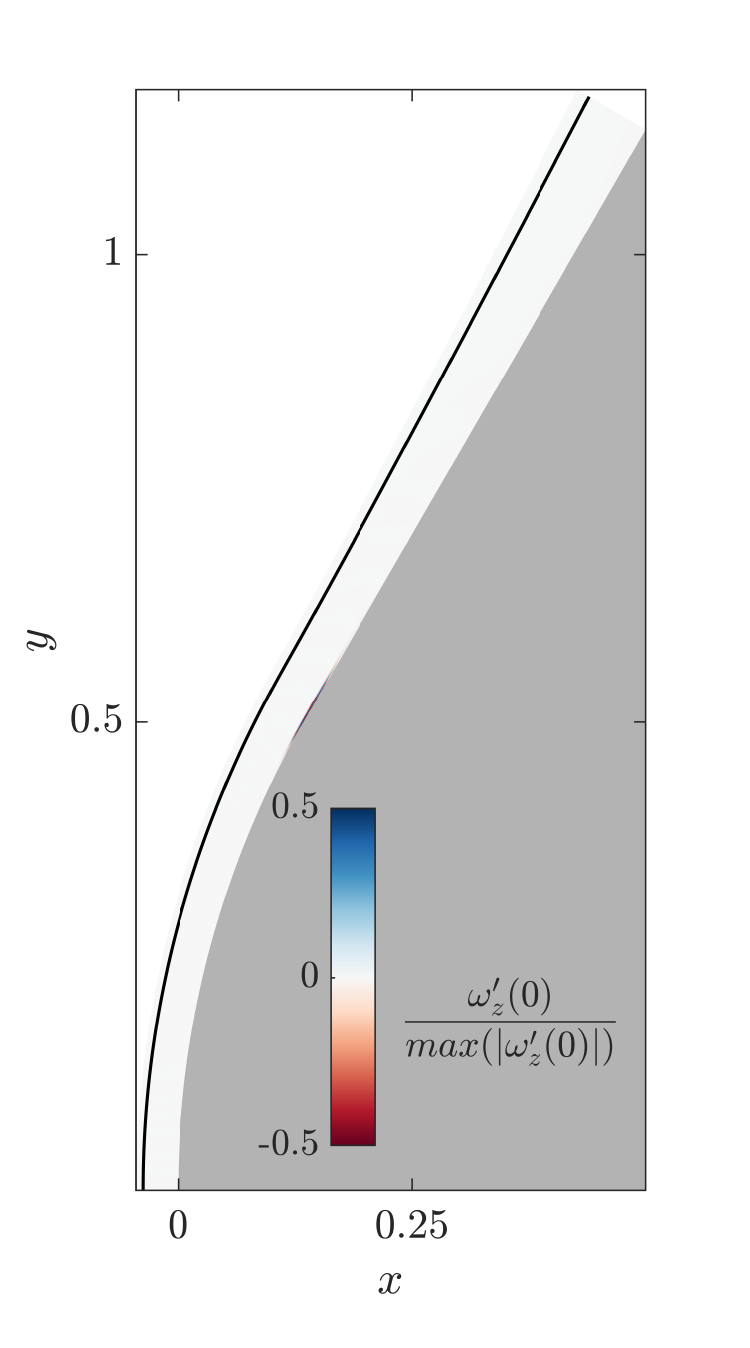}%
\end{minipage}%
}
    \subfloat[\label{fig:bl-branch-zoom}]{%
\begin{minipage}[t]{0.32\textwidth}
\centering
\includegraphics[width=\linewidth]{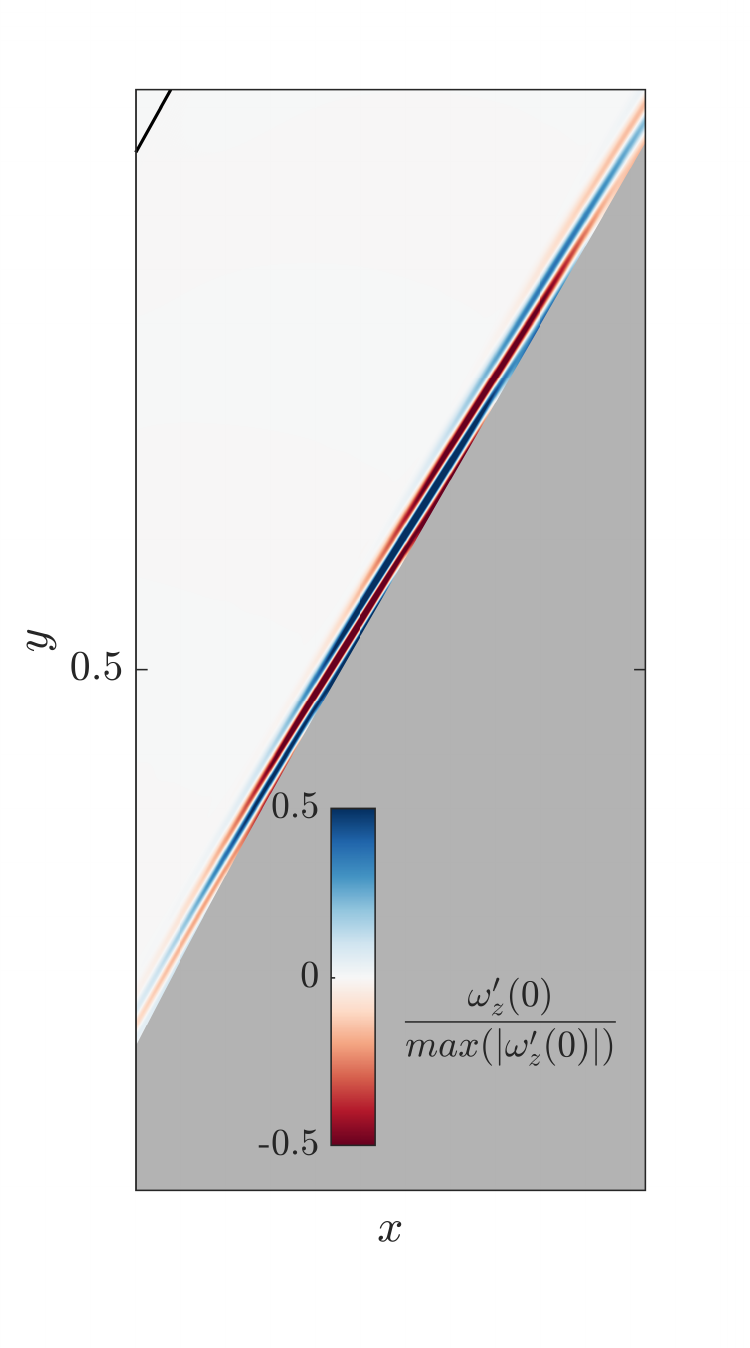}%
\end{minipage}%
}
    \caption{Optimal initial disturbance associated with the adiabatic
      boundary-layer-localized branch at $Re_\infty = 1\,000\,000$. The
      maximum gain on this branch is
      $G_{\max,\mathrm{BL}}^\text{opt}=480$. The base flow is computed
      from Table~\ref{tab:freestream} at $M_\infty =
      26.1$. (a) Disturbance vorticity. (b) Zoom of
      the near-wall structure.}
    \label{fig:boundary-layer-branch}
\end{figure}
\begin{figure}[tbp]
    \centering
    \subfloat[\label{fig:bl-branch-opt-vort}]{%
\begin{minipage}[t]{0.30\textwidth}
\centering
\includegraphics[width=\linewidth]{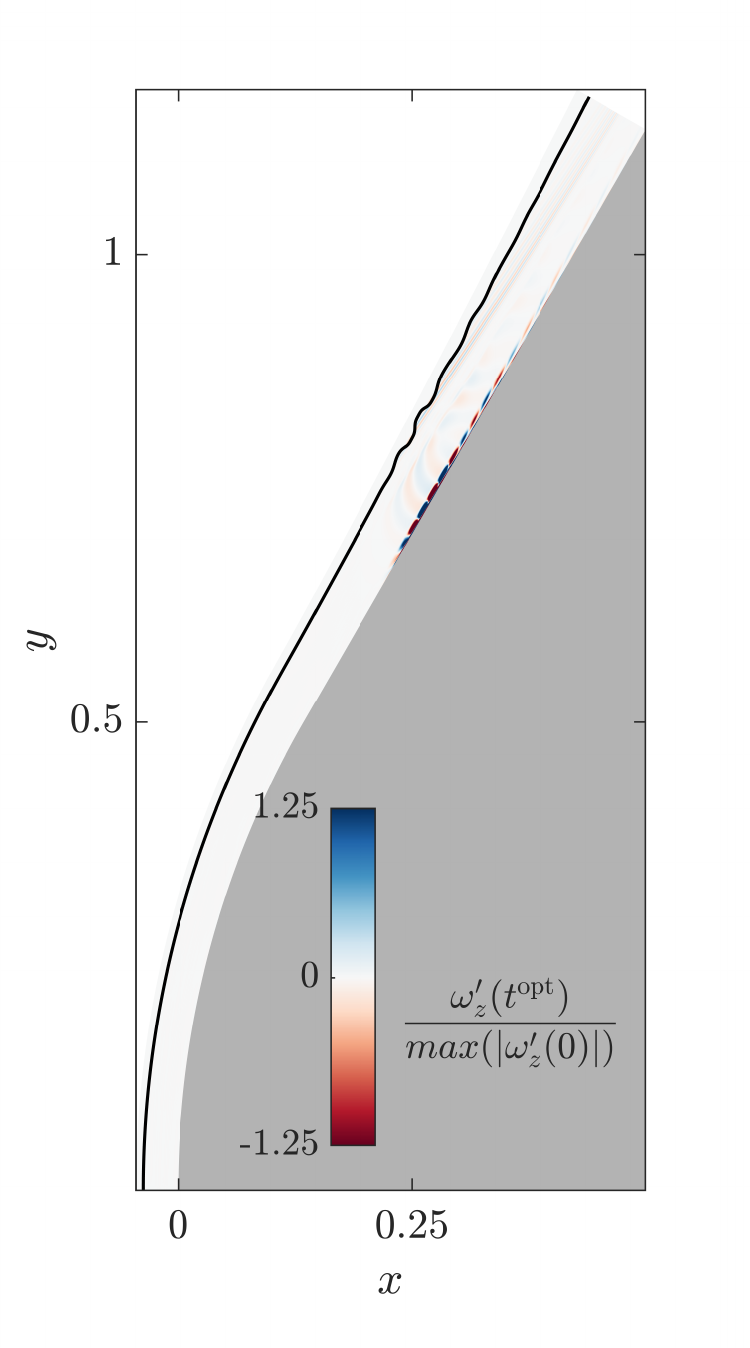}%
\end{minipage}%
}
    \subfloat[\label{fig:bl-branch-opt-vort-zoom}]{%
\begin{minipage}[t]{0.30\textwidth}
\centering
\includegraphics[width=\linewidth]{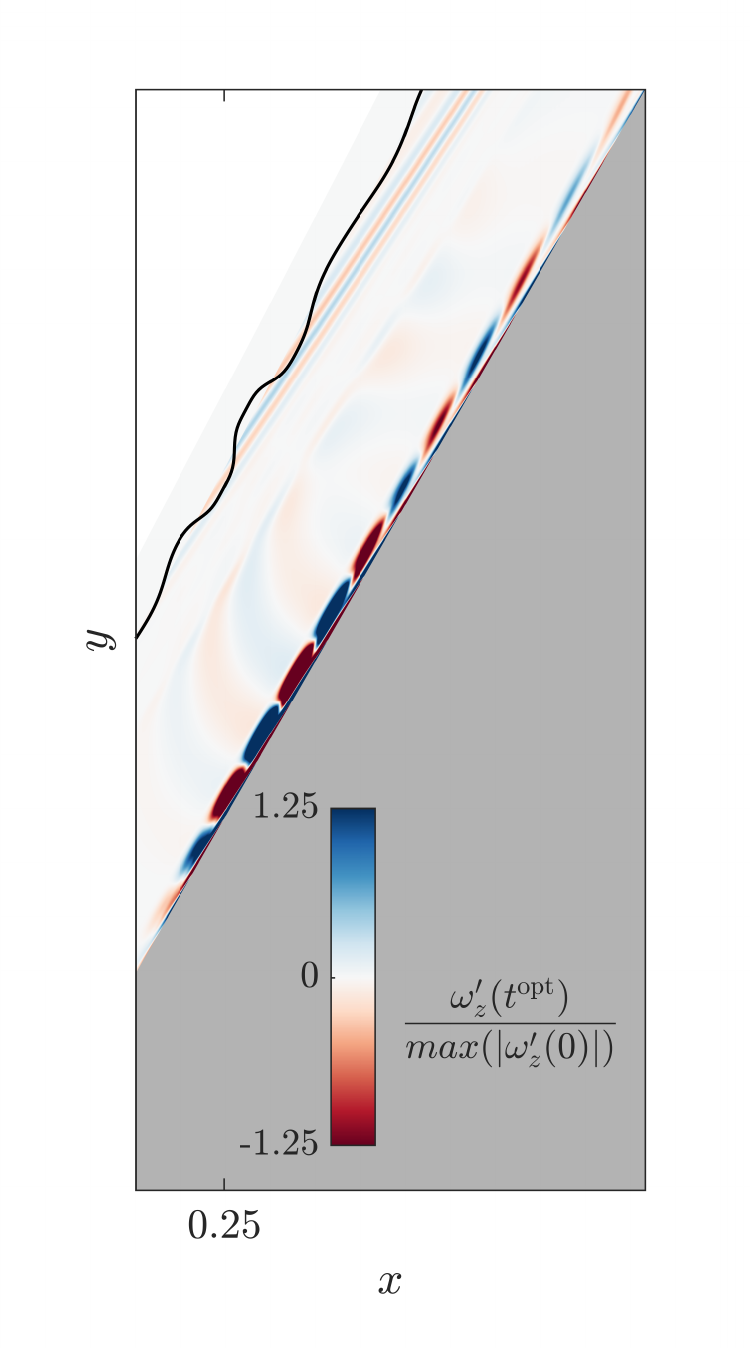}%
\end{minipage}%
}\\
    \subfloat[\label{fig:bl-branch-opt-div}]{%
\begin{minipage}[t]{0.30\textwidth}
\centering
\includegraphics[width=\linewidth]{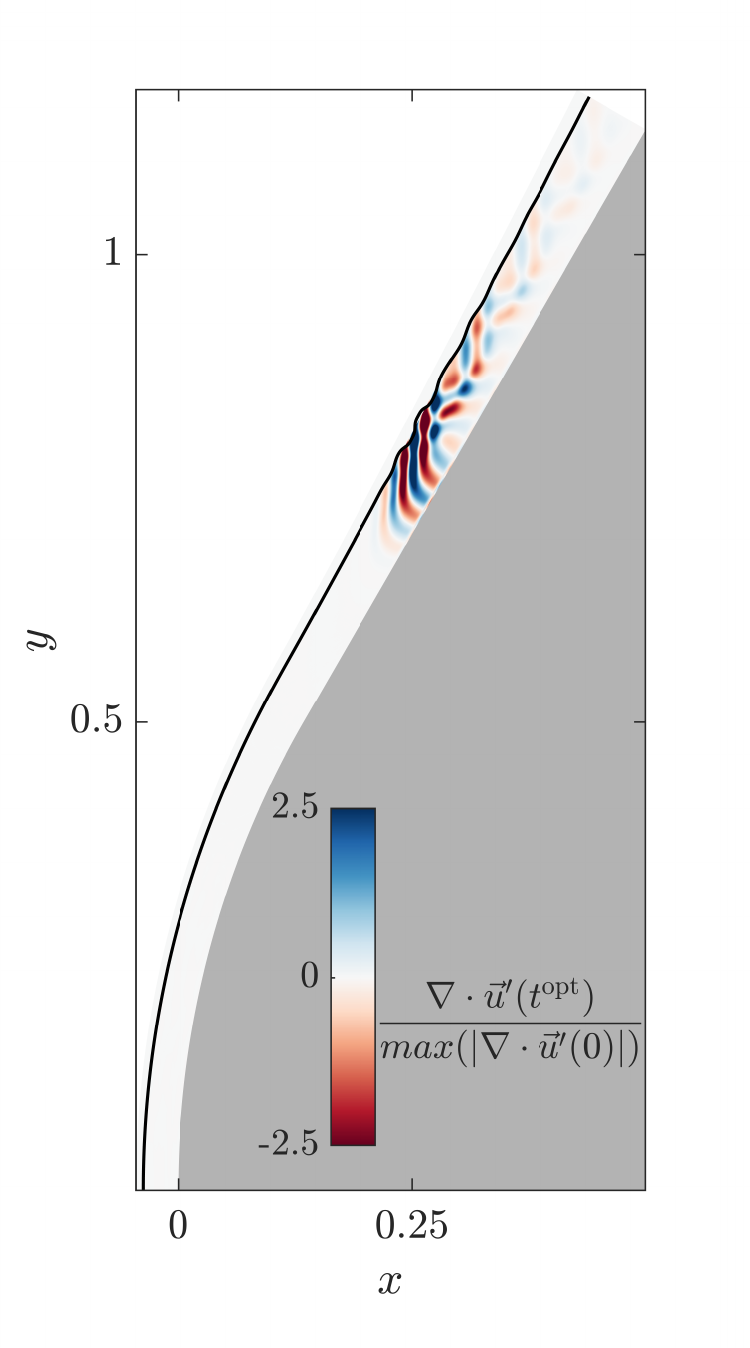}%
\end{minipage}%
}
    \subfloat[\label{fig:bl-branch-opt-div-zoom}]{%
\begin{minipage}[t]{0.30\textwidth}
\centering
\includegraphics[width=\linewidth]{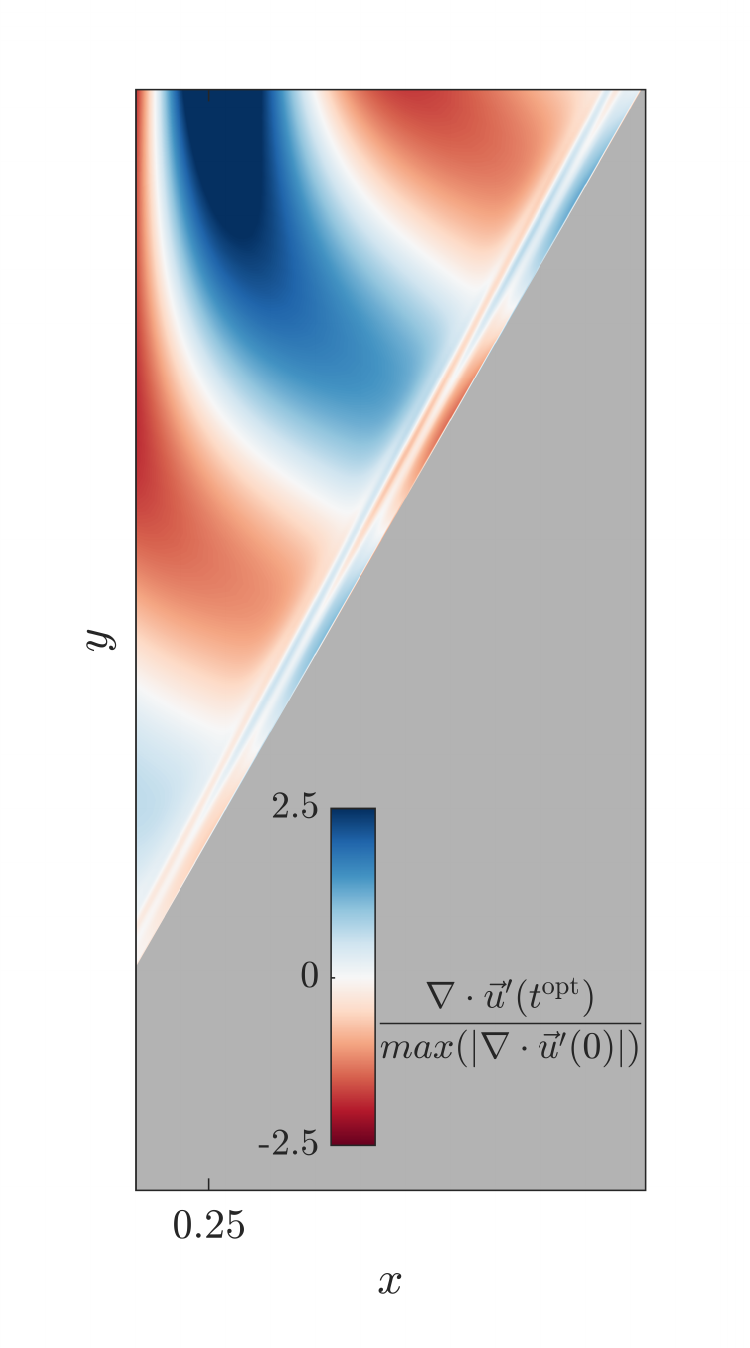}%
\end{minipage}%
}
    \caption{Optimal disturbance of the adiabatic
      boundary-layer-localized branch at its time of maximum
      amplification.  The maximum gain on this branch is
      $G_{\max,\mathrm{BL}}^\text{opt}=480$. Base flow computed from
      Table~\ref{tab:freestream} at $Re_\infty = 1\,000\,000$ and
      $M_\infty = 26.1$. (a) Disturbance vorticity.
      (b) Zoom of the disturbance-vorticity
      field. (c) Disturbance divergence.  (d) Zoom
      of the disturbance-divergence field.}
    \label{fig:boundary-layer-branch-opt-vort-div}
\end{figure}

At $Re_\infty \approx 10^6$, a second family of highly amplified
disturbances emerges within the wall boundary layer. Its structure is
shown in Fig.~\ref{fig:boundary-layer-branch}. These disturbances
are initially localized in the near-wall high-shear region rather than
in the outer entropy layer, yet their initial amplification mechanism
is analogous: the orientation of the velocity perturbations produces a
large positive Reynolds-stress work against the mean boundary-layer
shear, giving rise to vortical streaks that grow within the boundary
layer (see Fig.~\ref{fig:boundary-layer-branch-opt-vort-div}). As
these streaks intensify and impinge on the wall, they generate an
acoustic trapped mode confined within the boundary layer, as evidenced
by the divergence field in Fig.~\ref{fig:bl-branch-opt-div}. The
trapped acoustic mode radiates energy away from the boundary layer.
The outgoing acoustic waves reflect off the bow shock and return,
establishing a feedback loop between the shock and the boundary layer.
In addition, the resulting oscillations of the shock induce further
vortical streaks, as can be seen in
Fig.~\ref{fig:bl-branch-opt-vort-zoom}.
The appearance of this branch indicates that, at sufficiently high
Reynolds number, both the entropy layer and the wall boundary layer
can support strong transient growth. For the $O(10^5)$ regime
emphasized here, however, the entropy-layer branch remains dominant.
This regime is also relevant to Mars entry conditions: using the same
nose-radius-based definition of $Re_\infty$ adopted here, the
reconstructed peak-heating conditions reported for MSL and
Mars~2020/Perseverance imply $Re_\infty \approx 7\times 10^5$ and
$5\times 10^5$, respectively~\cite{bose2014reconstruction,
  edquist2022mars, alpert2022inverse}. These Reynolds-number estimates
suggest that entropy-layer transient growth may also be relevant
during the peak-heating phase of Martian entry of EDL vehicles.

\FloatBarrier
\section{Conclusions}
\label{sec:conclusions}

We have performed global modal and optimal transient-growth analyses
of the post-shock flow over a simplified axisymmetric blunt capsule
at a representative high-altitude Mars-entry condition. The aim was to
identify the dominant linear amplification mechanisms under
flight-relevant conditions for blunt EDL vehicles, for which the
classical picture of transition based on boundary-layer modal
instabilities remains incomplete.

The modal analysis found no temporally unstable axisymmetric global
modes over the Reynolds-number range examined. The real part of the
least stable eigenvalue remained negative and approached a finite
negative limit as $Re_\infty$ increased. Despite this modal stability,
the non-modal analysis revealed strong transient growth concentrated
in the shear--entropy layer generated by bow-shock curvature. The
energy budgets linked this growth to a two-stage amplification
process: Reynolds-stress production first extracted kinetic energy
from the mean shear, amplifying velocity fluctuations that
subsequently generated entropy fluctuations through their interaction
with the strong base-flow entropy gradient.

The optimal entropy-layer gain scales approximately linearly with
Reynolds number at sufficiently large $Re_\infty$, while a
boundary-layer-localized branch yields comparable gains near
$Re_\infty = 10^6$. Over the $O(10^5)$ Reynolds-number range relevant
to Mars entry, the entropy-layer mechanism remains dominant among
the axisymmetric disturbances considered here. These findings
suggest that shock-curvature-induced entropy-layer dynamics may
contribute to disturbance amplification preceding transition in
blunt entry vehicles. Accounting for this non-modal amplification
could complement classical boundary-layer modal analyses and help
inform transition prediction and aerothermal design.

\begin{acknowledgments}
The authors gratefully acknowledge Professor Hans G. Hornung for
valuable discussions on entropy-layer instabilities. We also thank
Professor Joseph E. Shepherd for generously sharing thermochemical
data files.

This work was supported by an Early Career Faculty grant from NASA's
Space Technology Research Grants Program (grant \#80NSSC23K1498).
\end{acknowledgments}

%

\appendix

\section{Validity of thermochemical equilibrium}
\label{sec:equilibrium-validity}

This appendix evaluates the thermochemical-equilibrium assumption
introduced in Sec.~\ref{sec:thermochem}. Chemical and
vibrational--electronic relaxation each possess a characteristic time
scale, and equilibrium is appropriate only when those time scales are
small compared with the relevant flow time scale. Because electronic
excitation becomes important only at temperatures above those
considered here, we restrict attention to chemical and vibrational
relaxation.

The chemical relaxation time $\tau_\text{chem}$ characterizes the rate at which 
translational–rotational energy relaxes toward its local thermochemical 
equilibrium value. Assuming first-order relaxation,
\begin{equation}
\frac{e_{T\text{-}R}(t)-e_{T\text{-}R}^{eq}}{e_{T\text{-}R}(0)-e_{T\text{-}R}^{eq}}
= \exp\!\left(-\frac{t}{\tau_\text{chem}}\right),
\end{equation}
so $\tau_\text{chem}$ can be extracted from the time $t_5$ at which the 
normalized departure from equilibrium has decayed to $5\%$:
\begin{equation}
\left.\frac{e_{T\text{-}R}(t)-e_{T\text{-}R}^{eq}}{e_{T\text{-}R}(0)-e_{T\text{-}R}^{eq}}\right|_{t=t_5} 
= 0.05 
\quad\Longrightarrow\quad
\tau_\text{chem}(P,T,X_i) = -\frac{t_5}{\ln(0.05)}.
\label{eq:tau-chem}
\end{equation}
In practice, $t_5$ is obtained by integrating an adiabatic, constant-pressure 
reactor in Cantera~\cite{cantera} from the prescribed initial state 
$(P,T,X_i)$ and recording the instant at which the $5\%$ threshold is first 
crossed.

For the vibrational relaxation time $\tau_\text{vib}$, we use the
Millikan--White
correlation~\cite{millikan1963systematics}. Estimating vibrational
relaxation for Mars mixtures is more difficult than for air because
the dominant species, $\mathrm{CO_2}$, possesses three vibrational
modes and four vibrational degrees of freedom. To obtain a
conservative estimate, we adopt the slowest mode---the asymmetric
stretch with $\theta = 3380$~K and $\mu_{\mathrm{CO_2\text{-}CO_2}} =
22$~g/mol. 

Both relaxation times require post-shock thermodynamic conditions
$(P,T,X_i)$ as input. Although those quantities vary along the shock,
we estimate them from the jump conditions across a normal
shock. Immediately behind the shock, only translational--rotational
modes are assumed to be excited~\cite{hill1986introduction}, and the
composition is taken to remain equal to the freestream composition
because the chemical time scale has not yet elapsed.

With these definitions, the Damk\"{o}hler numbers at $Re_\infty =
100\,000$ are $Da_\mathrm{chem}=O(10^2)$--$O(10^3)$ and
$Da_\mathrm{vib}=O(10^4)$--$O(10^5)$. Both are much larger than unity, which
confirms that chemical and vibrational relaxation are rapid relative
to the convective time scale. The thermochemical-equilibrium
assumption is therefore appropriate for the conditions studied here.

\section{Energy-budget terms}
\label{app:budget}

The local terms in Eqs.~\eqref{eq:kinetic-budget-main}
and~\eqref{eq:entropic-budget-main} are defined below, following
\citet{AntonAlvarez2026bow}. Repeated spatial indices imply summation.

For kinetic energy,
\begin{align}
\mathcal{A}^k
&= -\rho_0 u_{j,0}\frac{\partial}{\partial x_j}
   \left(\frac{u_i'u_i'}{2}\right), \notag\\
\mathcal{P}^k
&= -\rho_0 u_i'u_j'\frac{\partial u_{i,0}}{\partial x_j}
   -\rho' u_i'u_{j,0}\frac{\partial u_{i,0}}{\partial x_j}, \notag\\
\Pi_d^k
&= p'\frac{\partial u_i'}{\partial x_i}, \notag\\
\mathcal{T}^k
&= \frac{\partial}{\partial x_j}
   \left(-u_j'p'+u_i'\tau_{ij}'\right), \notag\\
\mathcal{D}^k
&= -\tau_{ij}'\frac{\partial u_i'}{\partial x_j}.
\label{eq:kinetic-budget-terms}
\end{align}

For entropic energy,
\begin{align}
\mathcal{A}^s
&= -\frac{(\gamma_0^*-1)p_0}{\gamma_0^*R_{g,0}^2}
   u_{j,0}\frac{\partial}{\partial x_j}
   \left(\frac{s'^2}{2}\right), \notag\\
\mathcal{P}^s
&= -\frac{(\gamma_0^*-1)p_0}{\gamma_0^*R_{g,0}^2}
   \left(u_j'+\frac{\rho'}{\rho_0}u_{j,0}
   +\frac{T'}{T_0}u_{j,0}\right)
   s'\frac{\partial s_0}{\partial x_j}, \notag\\
\mathcal{T}^s
&= -\frac{\gamma_0^*-1}{\gamma_0^*R_{g,0}}
   \frac{\partial(s'q_j')}{\partial x_j}, \notag\\
\mathcal{D}^s
&= \frac{\gamma_0^*-1}{\gamma_0^*R_{g,0}}
   q_j'\frac{\partial s'}{\partial x_j}, \notag\\
\mathcal{S}^s
&= \frac{\gamma_0^*-1}{\gamma_0^*R_{g,0}}s'\Phi'.
\label{eq:entropic-budget-terms}
\end{align}

Using Eqs.~\eqref{eq:ns-d} and~\eqref{eq:ns-e}, the stress,
heat-flux, and viscous-heating perturbations are
\begin{align}
\tau_{ij}'
&= \frac{\mu_0^*}{Re_\infty}
   \left(\frac{\partial u_i'}{\partial x_j}
   +\frac{\partial u_j'}{\partial x_i}
   -\frac{2}{3}\delta_{ij}\frac{\partial u_k'}{\partial x_k}\right)
   +\frac{\mu^{*\prime}}{\mu_0^*}\tau_{ij,0}, \notag\\
q_j'
&= -\frac{\gamma_\infty}{Re_\infty Pr_\infty}
   \left(k_0^*\frac{\partial T'}{\partial x_j}
   +k^{*\prime}\frac{\partial T_0}{\partial x_j}\right), \notag\\
\Phi'
&= \tau_{ij}'\frac{\partial u_{i,0}}{\partial x_j}
   +\tau_{ij,0}\frac{\partial u_i'}{\partial x_j}.
\label{eq:budget-flux-perturbations}
\end{align}
Here $\mu^{*\prime}$ and $k^{*\prime}$ include the linearized
thermodynamic dependence of the transport coefficients.
Domain-integrated terms are obtained by applying
$\langle\cdot\rangle_D$ to these local expressions.

\bibliography{paper}

@article{schneider2004hypersonic,
  author        = {Schneider, S. P.},
  title         = {Hypersonic laminar--turbulent transition on circular cones and scramjet forebodies},
  journal       = {Prog. Aerosp. Sci.},
  volume        = {40},
  number        = {1--2},
  pages         = {1--50},
  year          = {2004},
  doi           = {10.1016/j.paerosci.2003.11.001}
}

@article{zoby1981approximate,
  author        = {Zoby, E. V. and Moss, J. N. and Sutton, K.},
  title         = {Approximate convective-heating equations for hypersonic flows},
  journal       = {J. Spacecr. Rockets},
  volume        = {18},
  number        = {1},
  pages         = {64--70},
  year          = {1981},
  doi           = {10.2514/3.57788}
}

@inproceedings{hollis2005transition,
  author        = {Hollis, B. and Liechty, D. and Wright, M. and Holden, M. and Wadhams, T. and MacLean, M. and Dyakonov, A.},
  title         = {Transition onset and turbulent heating measurements for the {Mars Science Laboratory} entry vehicle},
  booktitle     = {43rd {AIAA} Aerospace Sciences Meeting and Exhibit},
  publisher     = {AIAA},
  year          = {2005},
  note          = {{AIAA} Paper 2005-1437},
  doi           = {10.2514/6.2005-1437}
}

@inproceedings{hollis2007turbulent,
  author    = {Hollis, B. R. and Collier, A. S.},
  title     = {Turbulent aeroheating testing of {Mars Science Laboratory} entry vehicle in perfect-gas nitrogen},
  booktitle = {45th {AIAA} Aerospace Sciences Meeting and Exhibit},
  publisher = {AIAA},
  year      = {2007},
  note      = {{AIAA} Paper 2007-1208},
  doi       = {10.2514/6.2007-1208}
}

@inproceedings{hollis2010blunt,
  author        = {Hollis, B. R.},
  title         = {Blunt-body entry vehicle aerothermodynamics: Transition and turbulence on the {CEV} and {MSL} configurations},
  booktitle     = {40th {AIAA} Fluid Dynamics Conference and Exhibit},
  publisher     = {AIAA},
  year          = {2010},
  note          = {{AIAA} Paper 2010-4984},
  doi           = {10.2514/6.2010-4984}
}

@article{wright2006modeling,
  author        = {Wright, M. J. and Olejniczak, J. and Brown, J. L. and Hornung, H. G. and Edquist, K. T.},
  title         = {Modeling of shock tunnel aeroheating data on the {Mars Science Laboratory} aeroshell},
  journal       = {J. Thermophys. Heat Transfer},
  volume        = {20},
  number        = {4},
  pages         = {641--651},
  year          = {2006},
  doi           = {10.2514/1.19896}
}

@article{bose2014reconstruction,
  author        = {Bose, D. and White, T. and Mahzari, M. and Edquist, K.},
  title         = {Reconstruction of aerothermal environment and heat shield response of {Mars Science Laboratory}},
  journal       = {J. Spacecr. Rockets},
  volume        = {51},
  number        = {4},
  pages         = {1174--1184},
  year          = {2014},
  doi           = {10.2514/1.A32783}
}

@inproceedings{edquist2022mars,
  author        = {Edquist, K. T. and Mahzari, M. and Alpert, H. S.},
  title         = {{Mars 2020} reconstructed aerothermal environments and design margins},
  booktitle     = {{AIAA} SciTech 2022 Forum},
  publisher     = {AIAA},
  year          = {2022},
  note          = {{AIAA} Paper 2022-0553},
  doi           = {10.2514/6.2022-0553}
}

@inproceedings{alpert2022inverse,
  author        = {Alpert, H. S. and Saunders, D. A. and Mahzari, M. and Monk, J. D. and White, T. R.},
  title         = {Inverse estimation of {Mars 2020} entry aeroheating environments using {MEDLI2} flight data},
  booktitle     = {{AIAA} SciTech 2022 Forum},
  publisher     = {AIAA},
  year          = {2022},
  note          = {{AIAA} Paper 2022-0550},
  doi           = {10.2514/6.2022-0550}
}

@techreport{lees1946investigation,
  author        = {Lees, L. and Lin, C. C.},
  title         = {Investigation of the stability of the laminar boundary layer in a compressible fluid},
  type          = {{NACA} Technical Note},
  number        = {1115},
  institution   = {National Advisory Committee for Aeronautics},
  year          = {1946},
  url           = {https://ntrs.nasa.gov/citations/19930081816}
}

@article{lin1945stability,
  author        = {Lin, C. C.},
  title         = {On the stability of two-dimensional parallel flows. {I. General} theory},
  journal       = {Q. Appl. Math.},
  volume        = {3},
  number        = {2},
  pages         = {117--142},
  year          = {1945},
  doi           = {10.1090/qam/13983}
}

@incollection{mack1984boundary,
  author        = {Mack, L. M.},
  title         = {Boundary-layer linear stability theory},
  booktitle     = {Special Course on Stability and Transition of Laminar Flow},
  series        = {{AGARD} Report},
  number        = {709},
  chapter       = {3},
  pages         = {3.1--3.81},
  publisher     = {Advisory Group for Aerospace Research and Development},
  year          = {1984},
  url           = {https://ntrs.nasa.gov/citations/19840025688}
}

@article{fedorov2011transition,
  author        = {Fedorov, A.},
  title         = {Transition and stability of high-speed boundary layers},
  journal       = {Annu. Rev. Fluid Mech.},
  volume        = {43},
  pages         = {79--95},
  year          = {2011},
  doi           = {10.1146/annurev-fluid-122109-160750}
}

@article{kimmel2015hifire,
  author        = {Kimmel, R. L. and Adamczak, D. and Paull, A. and Paull, R. and Shannon, J. and Pietsch, R. and Frost, M. and Alesi, H.},
  title         = {{HIFiRE-1} ascent-phase boundary-layer transition},
  journal       = {J. Spacecr. Rockets},
  volume        = {52},
  number        = {1},
  pages         = {217--230},
  year          = {2015},
  doi           = {10.2514/1.A32851}
}

@article{juliano2018hifire,
  author        = {Juliano, T. J. and Poggie, J. and Porter, K. M. and Kimmel, R. L. and Jewell, J. S. and Adamczak, D. W.},
  title         = {{HIFiRE-5b} heat flux and boundary-layer transition},
  journal       = {J. Spacecr. Rockets},
  volume        = {55},
  number        = {6},
  pages         = {1315--1328},
  year          = {2018},
  doi           = {10.2514/1.A34147}
}

@article{bertolotti1991analysis,
  author        = {Bertolotti, F. P. and Herbert, T.},
  title         = {Analysis of the linear stability of compressible boundary layers using the {PSE}},
  journal       = {Theor. Comput. Fluid Dyn.},
  volume        = {3},
  number        = {2},
  pages         = {117--124},
  year          = {1991},
  doi           = {10.1007/BF00271620}
}

@article{herbert1997parabolized,
  author        = {Herbert, T.},
  title         = {Parabolized stability equations},
  journal       = {Annu. Rev. Fluid Mech.},
  volume        = {29},
  pages         = {245--283},
  year          = {1997},
  doi           = {10.1146/annurev.fluid.29.1.245}
}

@article{bertolotti1992linear,
  author        = {Bertolotti, F. P. and Herbert, T. and Spalart, P. R.},
  title         = {Linear and nonlinear stability of the {Blasius} boundary layer},
  journal       = {J. Fluid Mech.},
  volume        = {242},
  pages         = {441--474},
  year          = {1992},
  doi           = {10.1017/S0022112092002453}
}

@techreport{chang1993linear,
  author        = {Chang, C.-L. and Malik, M. R. and Erlebacher, G. and Hussaini, M. Y.},
  title         = {Linear and nonlinear {PSE} for compressible boundary layers},
  type          = {{NASA} Contractor Report},
  number        = {191537},
  institution   = {NASA},
  year          = {1993},
  url           = {https://ntrs.nasa.gov/citations/19940015560}
}

@article{chang1994oblique,
  author        = {Chang, C.-L. and Malik, M. R.},
  title         = {Oblique-mode breakdown and secondary instability in supersonic boundary layers},
  journal       = {J. Fluid Mech.},
  volume        = {273},
  pages         = {323--360},
  year          = {1994},
  doi           = {10.1017/S0022112094001965}
}

@article{TowneColonius2015JCP,
  author        = {Towne, A. and Colonius, T.},
  title         = {One-way spatial integration of hyperbolic equations},
  journal       = {J. Comput. Phys.},
  volume        = {300},
  pages         = {844--861},
  year          = {2015},
  doi           = {10.1016/j.jcp.2015.08.015}
}

@article{TowneRigasPickeringColonius2022JFM,
  author        = {Towne, A. and Rigas, G. and Kamal, O. and Pickering, E. and Colonius, T.},
  title         = {Efficient global resolvent analysis via the one-way {Navier--Stokes} equations},
  journal       = {J. Fluid Mech.},
  volume        = {948},
  pages         = {A9},
  year          = {2022},
  doi           = {10.1017/jfm.2022.647}
}

@article{SleemanLakebrinkColonius2025AIAAJ,
  author        = {Sleeman, M. K. and Colonius, T. and Lakebrink, M. T.},
  title         = {Boundary-layer stability analysis using the nonlinear one-way {Navier--Stokes} approach},
  journal       = {AIAA J.},
  volume        = {63},
  number        = {8},
  pages         = {3145--3159},
  year          = {2025},
  doi           = {10.2514/1.J064909}
}

@article{Paredes2019,
  author        = {Paredes, P. and Choudhari, M. M. and Li, F.},
  title         = {Instability wave--streak interactions in a high {Mach} number boundary layer at flight conditions},
  journal       = {J. Fluid Mech.},
  volume        = {858},
  pages         = {474--499},
  year          = {2019},
  doi           = {10.1017/jfm.2018.744}
}

@article{Paredes2022,
  author        = {Paredes, P. and Scholten, A. and Choudhari, M. M. and Li, F. and Benitez, E. K. and Jewell, J. S.},
  title         = {Boundary-layer instabilities over a cone--cylinder--flare model at {Mach} 6},
  journal       = {AIAA J.},
  volume        = {60},
  number        = {10},
  pages         = {5652--5661},
  year          = {2022},
  doi           = {10.2514/1.J061829}
}

@article{theofilis2003advances,
  author        = {Theofilis, V.},
  title         = {Advances in global linear instability analysis of nonparallel and three-dimensional flows},
  journal       = {Prog. Aerosp. Sci.},
  volume        = {39},
  number        = {4},
  pages         = {249--315},
  year          = {2003},
  doi           = {10.1016/S0376-0421(02)00030-1}
}

@article{theofilis2011global,
  author        = {Theofilis, V.},
  title         = {Global linear instability},
  journal       = {Annu. Rev. Fluid Mech.},
  volume        = {43},
  pages         = {319--352},
  year          = {2011},
  doi           = {10.1146/annurev-fluid-122109-160705}
}

@article{paredes2016linear,
  author        = {Paredes, P. and Gosse, R. and Theofilis, V. and Kimmel, R.},
  title         = {Linear modal instabilities of hypersonic flow over an elliptic cone},
  journal       = {J. Fluid Mech.},
  volume        = {804},
  pages         = {442--466},
  year          = {2016},
  doi           = {10.1017/jfm.2016.536}
}

@article{hildebrand2018simulation,
  author        = {Hildebrand, N. and Dwivedi, A. and Nichols, J. W. and Jovanovi{\'c}, M. R. and Candler, G. V.},
  title         = {Simulation and stability analysis of oblique shock-wave/boundary-layer interactions at {Mach} 5.92},
  journal       = {Phys. Rev. Fluids},
  volume        = {3},
  number        = {1},
  pages         = {013906},
  year          = {2018},
  doi           = {10.1103/PhysRevFluids.3.013906}
}

@article{reshotko2001transient,
  author        = {Reshotko, E.},
  title         = {Transient growth: A factor in bypass transition},
  journal       = {Phys. Fluids},
  volume        = {13},
  number        = {5},
  pages         = {1067--1075},
  year          = {2001},
  doi           = {10.1063/1.1358308}
}

@article{schmid2007nonmodal,
  author        = {Schmid, P. J.},
  title         = {Nonmodal stability theory},
  journal       = {Annu. Rev. Fluid Mech.},
  volume        = {39},
  pages         = {129--162},
  year          = {2007},
  doi           = {10.1146/annurev.fluid.38.050304.092139}
}

@article{landahl1980note,
  author        = {Landahl, M. T.},
  title         = {A note on an algebraic instability of inviscid parallel shear flows},
  journal       = {J. Fluid Mech.},
  volume        = {98},
  number        = {2},
  pages         = {243--251},
  year          = {1980},
  doi           = {10.1017/S0022112080000122}
}

@article{butler1992three,
  author        = {Butler, K. M. and Farrell, B. F.},
  title         = {Three-dimensional optimal perturbations in viscous shear flow},
  journal       = {Phys. Fluids A},
  volume        = {4},
  number        = {8},
  pages         = {1637--1650},
  year          = {1992},
  doi           = {10.1063/1.858386}
}

@article{orr1907stability,
  author        = {Orr, W. {\relax M'F}},
  title         = {The stability or instability of the steady motions of a perfect liquid and of a viscous liquid. {Part II}: A viscous liquid},
  journal       = {Proc. R. Ir. Acad., Sect. A},
  volume        = {27},
  pages         = {69--138},
  year          = {1907},
  url           = {https://www.jstor.org/stable/20490591}
}

@article{hanifi1996transient,
  author        = {Hanifi, A. and Schmid, P. J. and Henningson, D. S.},
  title         = {Transient growth in compressible boundary layer flow},
  journal       = {Phys. Fluids},
  volume        = {8},
  number        = {3},
  pages         = {826--837},
  year          = {1996},
  doi           = {10.1063/1.868864}
}

@inproceedings{Bitter2014,
  author        = {Bitter, N. P. and Shepherd, J. E.},
  title         = {Transient growth in hypersonic boundary layers},
  booktitle     = {7th {AIAA} Theoretical Fluid Mechanics Conference},
  publisher     = {AIAA},
  year          = {2014},
  note          = {{AIAA} Paper 2014-2497},
  doi           = {10.2514/6.2014-2497}
}

@article{Paredes2016,
  author        = {Paredes, P. and Choudhari, M. M. and Li, F. and Chang, C.-L.},
  title         = {Optimal growth in hypersonic boundary layers},
  journal       = {AIAA J.},
  volume        = {54},
  number        = {10},
  pages         = {3050--3061},
  year          = {2016},
  doi           = {10.2514/1.J054912}
}

@article{Dwivedi2020,
  author        = {Dwivedi, A. and Hildebrand, N. and Nichols, J. W. and Candler, G. V. and Jovanovi{\'c}, M. R.},
  title         = {Transient growth analysis of oblique shock-wave/boundary-layer interactions at {Mach} 5.92},
  journal       = {Phys. Rev. Fluids},
  volume        = {5},
  number        = {6},
  pages         = {063904},
  year          = {2020},
  doi           = {10.1103/PhysRevFluids.5.063904}
}

@article{quintanilha2022transient,
  author        = {Quintanilha, Jr., H. and Paredes, P. and Hanifi, A. and Theofilis, V.},
  title         = {Transient growth analysis of hypersonic flow over an elliptic cone},
  journal       = {J. Fluid Mech.},
  volume        = {935},
  pages         = {A40},
  year          = {2022},
  doi           = {10.1017/jfm.2022.46}
}

@article{hornung2001shock,
  author        = {Hornung, H. G. and Lemieux, P.},
  title         = {Shock layer instability near the {Newtonian} limit of hypervelocity flows},
  journal       = {Phys. Fluids},
  volume        = {13},
  number        = {8},
  pages         = {2394--2402},
  year          = {2001},
  doi           = {10.1063/1.1383591}
}

@inproceedings{marineau2014mach,
  author        = {Marineau, E. C. and Moraru, G. C. and Lewis, D. R. and Norris, J. D. and Lafferty, J. F. and Wagnild, R. M. and Smith, J. A.},
  title         = {{Mach} 10 boundary layer transition experiments on sharp and blunted cones},
  booktitle     = {19th {AIAA} International Space Planes and Hypersonic Systems and Technologies Conference},
  publisher     = {AIAA},
  year          = {2014},
  note          = {{AIAA} Paper 2014-3108},
  doi           = {10.2514/6.2014-3108}
}

@article{jewell2017boundary,
  author        = {Jewell, J. S. and Kimmel, R. L.},
  title         = {Boundary-layer stability analysis for {Stetson's} {Mach} 6 blunt-cone experiments},
  journal       = {J. Spacecr. Rockets},
  volume        = {54},
  number        = {1},
  pages         = {258--265},
  year          = {2017},
  doi           = {10.2514/1.A33619}
}

@article{paredes2019nose,
  author        = {Paredes, P. and Choudhari, M. M. and Li, F. and Jewell, J. S. and Kimmel, R. L. and Marineau, E. C. and Grossir, G.},
  title         = {Nose-tip bluntness effects on transition at hypersonic speeds},
  journal       = {J. Spacecr. Rockets},
  volume        = {56},
  number        = {2},
  pages         = {369--387},
  year          = {2019},
  doi           = {10.2514/1.A34277}
}

@article{paredes2017blunt,
  author        = {Paredes, P. and Choudhari, M. M. and Li, F.},
  title         = {Blunt-body paradox and transient growth on a hypersonic spherical forebody},
  journal       = {Phys. Rev. Fluids},
  volume        = {2},
  number        = {5},
  pages         = {053903},
  year          = {2017},
  doi           = {10.1103/PhysRevFluids.2.053903}
}

@article{paredes2020mechanism,
  author        = {Paredes, P. and Choudhari, M. M. and Li, F.},
  title         = {Mechanism for frustum transition over blunt cones at hypersonic speeds},
  journal       = {J. Fluid Mech.},
  volume        = {894},
  pages         = {A22},
  year          = {2020},
  doi           = {10.1017/jfm.2020.261}
}

@article{scholten2024nonlinear,
  author        = {Scholten, A. and Paredes, P. and Choudhari, M. M. and Li, F. and Carpenter, M. and Bailey, M.},
  title         = {Nonlinear nonmodal analysis of hypersonic flow over blunt cones},
  journal       = {AIAA J.},
  volume        = {62},
  number        = {9},
  pages         = {3271--3283},
  year          = {2024},
  doi           = {10.2514/1.J063602}
}

@article{AntonAlvarez2026HYMOR,
  author        = {Ant{\'o}n-{\'A}lvarez, A. and Lozano-Dur{\'a}n, A.},
  title         = {{HYMOR}: An open-source package for modal, non-modal, and receptivity analysis in high-enthalpy hypersonic vehicles},
  journal       = {Comput. Phys. Commun.},
  volume        = {328},
  pages         = {110340},
  year          = {2026},
  doi           = {10.1016/j.cpc.2026.110340}
}

@article{poinsot1992boundary,
  author        = {Poinsot, T. J. and Lele, S. K.},
  title         = {Boundary conditions for direct simulations of compressible viscous flows},
  journal       = {J. Comput. Phys.},
  volume        = {101},
  number        = {1},
  pages         = {104--129},
  year          = {1992},
  doi           = {10.1016/0021-9991(92)90046-2}
}

@article{pandolfi2001numerical,
  author        = {Pandolfi, M. and D'Ambrosio, D.},
  title         = {Numerical instabilities in upwind methods: Analysis and cures for the ``carbuncle'' phenomenon},
  journal       = {J. Comput. Phys.},
  volume        = {166},
  number        = {2},
  pages         = {271--301},
  year          = {2001},
  doi           = {10.1006/jcph.2000.6652}
}

@article{chu1965energy,
  author        = {Chu, B.-T.},
  title         = {On the energy transfer to small disturbances in fluid flow ({Part I})},
  journal       = {Acta Mech.},
  volume        = {1},
  number        = {3},
  pages         = {215--234},
  year          = {1965},
  doi           = {10.1007/BF01387235}
}

@misc{cantera,
  author        = {Goodwin, D. G. and Moffat, H. K. and Schoegl, I. and Speth, R. L. and Weber, B. W.},
  title         = {{Cantera}: An object-oriented software toolkit for chemical kinetics, thermodynamics, and transport processes},
  howpublished  = {\url{https://www.cantera.org}},
  year          = {2024},
  note          = {{Version} 3.1.0},
  doi           = {10.5281/zenodo.14455267}
}

@article{millikan1963systematics,
  author        = {Millikan, R. C. and White, D. R.},
  title         = {Systematics of vibrational relaxation},
  journal       = {J. Chem. Phys.},
  volume        = {39},
  number        = {12},
  pages         = {3209--3213},
  year          = {1963},
  doi           = {10.1063/1.1734182}
}

@book{hill1986introduction,
  author        = {Hill, T. L.},
  title         = {An Introduction to Statistical Thermodynamics},
  publisher     = {Dover},
  address       = {New York},
  year          = {1986}
}

@misc{AntonAlvarez2026bow,
  author        = {Ant{\'o}n-{\'A}lvarez, A. and Lozano-Dur{\'a}n, A.},
  title         = {Bow-shock instability in entry, descent, and landing vehicles under high-enthalpy conditions},
  year          = {2026},
  eprint        = {2605.28357},
  archivePrefix = {arXiv},
  primaryClass  = {physics.flu-dyn},
  doi           = {10.48550/arXiv.2605.28357}
}

\end{document}